\documentclass[aps,prb,reprint,superscriptaddress,longbibliography]{revtex4-2} 

\usepackage{amsmath,amssymb,bm,braket}
\usepackage{enumitem}

\usepackage{amsthm}
\newtheorem{theorem}{Theorem}
\newtheorem{proposition}{Proposition}
\newtheorem{lemma}{Lemma}
\newtheorem*{main}{Main Result}

\usepackage{xcolor}

\usepackage{graphicx}
\graphicspath{{./}{./figs/}}

\usepackage{array}
\usepackage{colortbl}
\newcommand{\Ic}[1]{\multicolumn{1}{|c}{#1}}
\newcommand{\cI}[1]{\multicolumn{1}{c|}{#1}}

\newcommand{\IIc}[1]{\multicolumn{1}{||c}{#1}}

\newcommand{\IcII}[1]{\multicolumn{1}{|c||}{#1}}
\newcommand{\IIcII}[1]{\multicolumn{1}{||c||}{#1}}
\newcommand{\ucline}[1]{\cline{#1}\noalign{\vspace{-0.8pt}}\cline{#1}}
\newcommand{\dcline}[1]{\cline{#1}\noalign{\vspace{-0.8pt}}\cline{#1}}
\newcommand{\gray}[1]{\multicolumn{1}{>{\columncolor[gray]{0.8}[4pt]}c}{#1}}
\newcommand{\Igray}[1]{\multicolumn{1}{|>{\columncolor[gray]{0.8}[4pt]}c}{#1}}

\newcommand{\IgrayI}[1]{\multicolumn{1}{|>{\columncolor[gray]{0.8}[4pt]}c|}{#1}}

\usepackage{hyperref}
\hypersetup{bookmarksnumbered=true,bookmarksdepth=subsubsection,breaklinks=true,colorlinks=true, citecolor=magenta, linkcolor=blue, urlcolor=cyan}

\begin{document}
\setlength{\abovedisplayskip}{8pt}%
\setlength{\belowdisplayskip}{8pt}%

\title{Proof of absence of local conserved quantities in two- and higher-dimensional quantum Ising models}

\author{Yuuya Chiba}
\email{yuya.chiba@riken.jp}
\affiliation{Nonequilibrium Quantum Statistical Mechanics RIKEN Hakubi Research Team, Pioneering Research Institute (PRI), RIKEN, 2-1 Hirosawa, Wako, Saitama 351-0198, Japan}

\date{\today}

\begin{abstract}
We prove that the Ising models with transverse and longitudinal fields on the hypercubic lattices with dimensions higher than one have no local conserved quantities other than the Hamiltonian.  This holds for any value of the longitudinal field, including zero, as far as the transverse field and the Ising interactions are nonzero.  The conserved quantity considered here is ``local'' in a very weak sense: it can be written as a linear combination of operators whose side lengths of the supports in one direction do not exceed half the system size, while the side lengths in the other directions are arbitrary.  We also prove that the above result holds even in the ladder system.  Our results extend the recently developed technique of the proof of absence of local conserved quantities in one-dimensional systems to higher dimensions and to the ladder.
\end{abstract}




\maketitle

\section{\label{sec:Introduction}Introduction}

Statistical mechanics describes thermal equilibrium states in macroscopic systems,
usually by utilizing a finite number of local conserved quantities,
such as energy and particle numbers~\cite{Landau1980,Callen1985}.
As suggested from the studies of thermalization in isolated quantum many-body systems~\cite{DAlessio2016,Mori2018},
if some local conserved quantities are neglected,
then the state to which the system relaxes will not be correctly described.
In particular, it is known that integrable systems, 
whose number of local conserved quantities becomes arbitrarily large with respect to the system size,
do not thermalize, i.e., do not relax to the thermal equilibrium state~\cite{Rigol2007,Rigol2008}.

In addition to this ordinary violation of thermalization, further studies have discovered many other violations of thermalization, such as quantum many-body scars~\cite{Bernien2017,Turner2018,Turner2018a,Moudgalya2018,Moudgalya2018a} and the Hilbert space fragmentations~\cite{DeTomasi2019,Moudgalya2021}.
These phenomena have been found to occur even 
on lattices other than chains~\cite{Lin2020,VanVoorden2020,Michailidis2020,Ren2022,Dai2024,Yoshinaga2022,Khudorozhkov2022,Will2024}, and extensive studies including several experiments~\cite{Bluvstein2021,Yao2023,Adler2024} have been devoted to elucidating such phenomena 
on various lattices.
Furthermore, in higher-dimensional systems,
the spontaneous symmetry breaking can occur 
at a finite temperature,
and it significantly affects
thermalization phenomena~\cite{Fratus2015,Mondaini2016,Blass2016,Mondaini2017,Richter2020}.
In particular, it has been shown that 
for a certain naive quench process 
in the two-dimensional Ising model with a magnetic field,
thermalization does not occur
due to the symmetry breaking in the prequench system~\cite{Reimann2021}.
All these violations of thermalization, 
namely quantum many-body scars, the Hilbert space fragmentations, and phenomena related to spontaneous symmetry breaking, 
are expected to
be independent of 
the ordinary violations 
caused by local conserved quantities (or integrability),
as suggested by several numerical calculations~\cite{Mondaini2016,Noh2023}, 
such as those of energy level spacing distributions~\cite{Mehta2004,Atas2013}.
To guarantee this independence theoretically,
it would be desirable if we could prove that these systems have no nontrivial local conserved quantities.

The proof of absence of local conserved quantities 
has long been out of the scope of theoretical analysis.
Recently, inspired by the pioneering result by Shiraishi~\cite{Shiraishi2019}, 
such proofs have been obtained in many one-dimensional systems~\cite{Shiraishi2019,Chiba2024,Park2024,Shiraishi2024,Chiba2024a,Park2024a,Yamaguchi2024,Yamaguchi2024a,Hokkyo2024a}.
However, the proofs in two- and higher-dimensional systems remain lacking.

In this paper, we prove absence of local conserved quantities in the Ising model with transverse and longitudinal fields on two- and higher-dimensional hypercubic lattices.
Our proof applies to both zero and nonzero longitudinal field cases, as far as the transverse field and the Ising interactions are nonzero.
The ``local'' conserved quantities we examine here are in fact conserved quantities satisfying a very weak condition on its locality:
there is at least one spatial direction such that 
the conserved quantity can be written as a linear combination of operators, 
each of which is supported on some region 
whose side length in that direction is less than or equal to half the system size.
This means that the side lengths in any other directions are arbitrary.
We show that such a conserved quantity is restricted only to the Hamiltonian, i.e., the trivial one,
for those quantum Ising models.
Furthermore, we prove that the same result holds even in the ladder case (an intermediate case between one and two dimensions), 
where the above locality condition reduces to 
almost the same form as the ordinary one in one-dimensional systems.

The paper is organized as follows.
Section~\ref{sec:Setup} describes the model, the definition of local conserved quantities, and our main result.
Section~\ref{sec:Integrability} explains how our main result is related to quantum nonintegrability and how it is consistent with the existing numerical studies.
Section~\ref{sec:Outline} introduces expressions of local conserved quantities in terms of the Pauli product basis.
The proof of the main result in the model on the two-dimensional square lattice is given in Sec.~\ref{sec:2D}, which is the main part of this paper. 
The extension to systems with dimensions higher than two is given in Sec.~\ref{sec:3D}
and to the ladder system in Sec.~\ref{sec:Ladder}.
Section~\ref{sec:ExtensionOtherLattice} discusses potential extension to other types of two-dimensional lattices.
Section~\ref{sec:Discussion} concludes the paper.

\section{\label{sec:Setup}Setup and main result}

Let $\Lambda=\{1,...,L\}^d$ be a $d$-dimensional hypercubic lattice with side length $L$ and $\vec{e}_{\mu}$ be its primitive translation vector in the $\mu$th direction ($\mu=1,...,d$).
The Hamiltonian of the quantum spin-$1/2$ Ising model on $\Lambda$ is given by
\begin{align}
    H=\sum_{\vec{r}\in\Lambda}\Bigl(\sum_{\mu=1}^{d}J^{\mu}Z_{\vec{r}}Z_{\vec{r}+\vec{e}_{\mu}}+h^{x}X_{\vec{r}}+h^{z}Z_{\vec{r}}\Bigr),
    \label{eq:Hamiltonian}
\end{align}
where $X_{\vec{r}},Y_{\vec{r}},Z_{\vec{r}}$ are the Pauli operators on site $\vec{r}$, 
$J^{\mu}$ is the Ising interaction between neighboring sites $\vec{r}$ and $\vec{r}+\vec{e}_{\mu}$, 
and $h^{x}$ ($h^{z}$) is the transverse (longitudinal) field.
Here we impose the periodic boundary conditions.

To describe our main result, we give a precise definition of
local conserved quantities.
We say that an operator is an $(\ell_{1},...,\ell_{d})$-support operator 
if the smallest rectangular cuboid containing its support has side lengths $\ell_{1},...,\ell_{d}$.
For instance, 
$Z_{\vec{r}}Z_{\vec{r}+\vec{e}_{1}}$ is a $(2,1,...,1)$-support operator.
Using this, we say that an operator commuting with $H$ is a $(k_{1},...,k_{d})$-local conserved quantity
if it is a sum of $(\ell_{1},...,\ell_{d})$-support operators where the maximum values of $\ell_{1},...,\ell_{d}$ appearing in the sum are given by $k_{1},...,k_{d}$, respectively.
For instance, the Hamiltonian $H$ itself is a $(2,...,2)$-local conserved quantity but we do not say that it is a $(3,...,3)$-local conserved quantity. 
For a detailed expression of the sum in the definition of $(k_{1},...,k_{d})$-local conserved quantities, see Eq.~\eqref{eq:Q_Expansion}.

Using these notions, our main result can be expressed as follows. 
\begin{main}
Suppose that 
the coupling constants in model~\eqref{eq:Hamiltonian} other than $h^z$ are nonzero 
and that 
the lattice dimension~$d$ satisfies $d\ge 2$. 
Let $k_{1},...,k_{d}$ be integers in $\{1,...,L\}$ and assume that one of them (say $k_{\mu^*}$) satisfies $k_{\mu^*}\le L/2$. When $3\le k_{\mu^*}\le L/2$, the model has no $(k_1,...,k_d)$-local conserved quantity. 
Furthermore, when $k_{\mu^*}\le 2$, any $(k_1,...,k_d)$-local conserved quantity is restricted to a linear combination of the Hamiltonian~$H$ and the identity~$I$ [i.e., a $(2,...,2)$-local one].
\end{main}

Since the above result surely applies to any $k_{1},...,k_{d}$ satisfying $k_{1},...,k_{d}\ll L$, it proves absence of ``usual'' local conserved quantities, which correspond to $(k_{1},...,k_{d})$-local conserved quantities with $k_{1},...,k_{d}=O(L^0)$ in our terminology.
Furthermore, it also proves absence of certain highly nonlocal conserved quantities, such as $(3, L, L,..., L)$-local ones and $(L/2, L, L,..., L)$-local ones, because $k_{1},...,k_{d}$ other than one of them (denoted by $k_{\mu^*}$) can be taken arbitrarily large. 
This seems surprising because one might naively expect that not just one of $k_{1},...,k_{d}$ but all of them must be sufficiently small.

Note that while the existing proofs of absence of local conserved quantities~\cite{Shiraishi2019,Chiba2024,Park2024,Shiraishi2024,Chiba2024a,Park2024a,Yamaguchi2024,Yamaguchi2024a,Hokkyo2024a} have been carried out in one-dimensional systems, our result treats two- and higher-dimensional systems.
Thus, our result 
will be the first proof of absence of local conserved quantities in systems with dimensions higher than one.

Note also that the above result is applicable to the transverse-field Ising model ($h^{z}=0$ case). 
This fact is contrasted with the previous proof of absence of local conserved quantities in the one-dimensional case~\cite{Chiba2024}, where $h^{z}\neq 0$ is necessary.

\section{\label{sec:Integrability}Relation to quantum integrability}

Here we give some comments on relation to quantum (non)integrability.
Quantum (non)integrability in a many-body-physics context is often characterized by using local conserved quantities
and is discussed in relation to the constructability of all energy eigenstates.
Although the definition of quantum (non)integrability has not been established~\cite{Caux2011,Gogolin2016,Mori2018}, it is usually considered that in any quantum integrable systems (described by a local and translation-invariant Hamiltonian) 
the number of ``usual'' local conserved quantities [$(k_{1},...,k_{d})$-local ones with $k_{1},...,k_{d}=O(L^0)$] becomes infinite after the limit $L\to\infty$.
Indeed, 
in one-dimensional integrable systems solvable by the algebraic Bethe ansatz, 
$k_{1}$-local conserved quantities are produced from the transfer matrix for any $k_{1}\lesssim L/2$~\cite{Sklyanin1979,Takhtadzhan1979,Korepin1993}, 
and in free fermionic integrable systems, 
there are $(k_{1},...,k_{d})$-local conserved quantities that can be written as quadratic forms of fermionic operators for any $k_{1},...,k_{d}\lesssim L/2$~\footnote{It is straightforward to extend the expressions of local conserved quantities of translation-invariant free fermionic systems, which is obtained for the one-dimensional case in Appendix~C of Ref.~\cite{Fagotti2013} (See also Refs.~\cite{Fagotti2014,Fagotti2016}.), 
to the higher dimensional cases.}.
By adopting this, i.e., assuming that for all integrable systems 
the number of usual local conserved quantities becomes arbitrarily large~\footnote{%
Note that the opposite is not true, that is, 
the existence of 
an arbitrarily large
number of usual local conserved quantities does not always imply integrability,
as exemplified by some model that exhibits quantum chaotic behaviors 
(thus constructing all energy eigenstates seems impossible) 
but its number of usual local conserved quantities becomes arbitrarily large~\cite{Hamazaki2016}.
},
we can say that a model possessing only a small number of usual local conserved quantities is nonintegrable.
Under this assumption, the main result given in the previous section can be regarded as a proof of nonintegrability.

In particular, since 
our main result
holds regardless of whether $h^{z}$ is zero or not, 
it indicates that there are no quantum integrable systems in the parameter space $(J^{1},..., J^{d},h^{x},h^{z})$ with $J^{1},..., J^{d},h^{x}\neq 0$ (including $h^{z}=0$). 
By contrast, the $h^{x}=0$ case is trivially quantum integrable~\footnote{Although thermodynamic functions of the $h^{x}=0$ case have been calculated analytically only for $h^{z}= 0$ and $d= 2$, such a fact is not related to quantum integrability, and all the $h^{x}=0$ cases, including $h^{z}\neq 0$ or $d>2$, are quantum integrable systems.}.
Moreover, if some of $J^{1},..., J^{d}$ are zero, the Hamiltonian is written as an array of lower-dimensional systems, and hence the analysis reduces to a smaller-$d$ case.
In addition, in the one-dimensional model, 
all values of parameters have been completely classified into either integrable or nonintegrable~\cite{Chiba2024}.
Therefore, the present results, combined with the above discussion, complete the classification of integrability and nonintegrability for all values of parameters $(J^{1},..., J^{d},h^{x},h^{z})$ and dimension $d$.

Absence of local conserved quantities, or nonintegrability, was also investigated numerically.
When numerically testing whether a given quantum many-body system is nonintegrable, one usually examines the energy level spacing distribution~\cite{Mehta2004,Atas2013}.
It is usually said that this distribution should be constructed using energy eigenvalues taken from a subspace specified by quantum numbers of all discrete unitary symmetries~\footnote{Note that the definition of such symmetries is sometimes the subject of debate because if one allowed arbitrary unitaries as candidates of unitary symmetries, any Hamiltonian on a $D$-dimensional Hilbert space would have at least $D$ number of independent unitary symmetries.} and all local conserved quantities, such as the translation symmetry and the total magnetization.
If there remain any other unitary symmetries or local conserved quantities, then level repulsion will be absent between energy eigenvalues corresponding to different quantum numbers of the remaining symmetries or local conserved quantities.
In other words, if the distribution is well described by the Wigner-Dyson distribution, which exhibits level repulsion, it suggests that there will be neither other unitary symmetries nor other local conserved quantities.

By examining such a level spacing distribution, Ref.~\cite{Mondaini2016} numerically shows that the two-dimensional case ($d=2$) of model~\eqref{eq:Hamiltonian} with certain values of parameters, such as $(J^{1},J^{2},h^x,h^z)=(1,1,1,0)$ and $(-1,-1,1,1)$, does not possess local conserved quantities other than the Hamiltonian, which is consistent with our main result.
However, in addition to the fact that such a numerical approach is not a rigorous proof, it contains two difficulties.
One difficulty is that each numerical result can tell only about each choice of values of parameters, and thus the whole parameter space cannot be covered by a finite number of numerical calculations. 
This means that we cannot exclude the possibility of the existence of unknown integrable points in the parameter space.
The other difficulty is that such a numerical approach suffers from a finite size effect, especially near the integrable points.
Therefore, it would be sometimes considered controversial whether the integrability of model~\eqref{eq:Hamiltonian} at $h^x=0$ persists at some small but nonzero $h^x$~\cite{Prosen2000,Brandino2015,Yu2022,Michailidis2020a}.
In contrast to such a numerical approach, our result rigorously shows that there is no unknown integrable system in the parameter space of model~\eqref{eq:Hamiltonian}, and the model is nonintegrable even at arbitrarily small but nonzero $h^x$.

Note that, although our main result can be applied to the two-dimensional transverse-field Ising model 
(the case of $d=2$ and $h^{z}=0$) with an arbitrarily small $h^{x}\neq 0$, it is not applicable to the effective Hamiltonian obtained by taking a certain weak-transverse-field limit~\cite{Yoshinaga2022}.
This can be seen from the fact that 
the effective Hamiltonian conserves the domain wall number
while the Hamiltonian~\eqref{eq:Hamiltonian} does not.
It is known that the early dynamics by the true $H$ for small $h^{x}$ can be well described by the effective Hamiltonian up to the timescale where prethermalization ends~\cite{Yoshinaga2022}. 
These facts demonstrate that the effect of presence or absence of local conserved quantities is crucial at a sufficiently large time (after the prethermalization) rather than an early time.

\section{\label{sec:Outline}Expression of candidates of local conserved quantities}

In order to represent a candidate of a $(k_{1},...,k_{d})$-local conserved quantity explicitly, we use Pauli products as an operator basis. Let $R^{(\ell_{1},...,\ell_{d})}_{\vec{r}}\subset\Lambda$ be a rectangular cuboid defined by
\begin{align}
    R^{(\ell_{1},...,\ell_{d})}_{\vec{r}}=\prod_{\mu=1}^{d}\{r_{\mu},...,r_{\mu}+\ell_{\mu}-1\},
\end{align}
where $\vec{r}=(r_{1},...,r_{d})$ is one of its corners and $\ell_{\mu}$ is its side length in the $\mu$th direction. 
Let $\bm{A}_{\vec{r}}^{(\ell_{1},...,\ell_{d})}$ denotes a Pauli product where the smallest rectangular cuboid containing its support is given by $R^{(\ell_{1},...,\ell_{d})}_{\vec{r}}$.
It is expressed as
\begin{align}
    \bm{A}_{\vec{r}}^{(\ell_{1},...,\ell_{d})}=\prod_{\vec{p}\in R^{(\ell_{1},...,\ell_{d})}_{\vec{r}}} A_{\vec{p}},
    \label{eq:PauliProd_A}
\end{align}
where $A_{\vec{p}}=X_{\vec{p}},Y_{\vec{p}},Z_{\vec{p}},I_{\vec{p}}$. Since $R^{(\ell_{1},...,\ell_{d})}_{\vec{r}}$ is the smallest rectangular cuboid, 
there is a Pauli operator $A_{\vec{p}}\neq I_{\vec{p}}$ on every face of $R^{(\ell_{1},...,\ell_{d})}_{\vec{r}}$.

Using these, a candidate of a $(k_{1},...,k_{d})$-local conserved quantity $Q$ 
(without loss of generality, $Q$ can be taken traceless)
is represented as
\begin{align}
    Q=\sum_{\vec{r}\in\Lambda}\sum_{\ell_{1}=1}^{k_{1}}...\sum_{\ell_{d}=1}^{k_{d}}\sideset{}{_{(\vec{r})}^{(\ell_{1},...,\ell_{d})}}{\sum}_{\bm{A}}
    c_{\bm{A},\vec{r}}^{(\ell_{1},...,\ell_{d})}
    \bm{A}_{\vec{r}}^{(\ell_{1},...,\ell_{d})}.
    \label{eq:Q_Expansion}
\end{align}
Here $\bm{A}_{\vec{r}}^{(\ell_{1},...,\ell_{d})}$ represents a Pauli product on $R^{(\ell_{1},...,\ell_{d})}_{\vec{r}}$ explained above Eq.~\eqref{eq:PauliProd_A}, and $\bm{A}$ in the sum~$\displaystyle \sideset{}{_{(\vec{r})}^{(\ell_{1},...,\ell_{d})}}{\sum}_{\bm{A}}$ runs over all such Pauli products.
By definition, any $(k_{1},...,k_{d})$-local conserved quantity must include, for any $\mu=1,...,d$, some nonzero expansion coefficient $c_{\bm{A},\vec{r}}^{(\ell_{1},...,\ell_{d})}$ with $\ell_{\mu}=k_{\mu}$.

Furthermore, since $H$ includes only onsite and nearest-neighbor terms, we can also expand the commutator $[Q, H]$ 
as 
\begin{align}
    \frac{[Q,H]}{2i}=\sum_{\vec{r}\in\Lambda}\sum_{\ell_{1}=1}^{k_{1}+1}...\sum_{\ell_{d}=1}^{k_{d}+1}\sideset{}{_{(\vec{r})}^{(\ell_{1},...,\ell_{d})}}{\sum}_{\bm{B}}
    r_{\bm{B},\vec{r}}^{(\ell_{1},...,\ell_{d})}
    \bm{B}_{\vec{r}}^{(\ell_{1},...,\ell_{d})},
    \label{eq:[Q,H]_Expansion}
\end{align}
where $\bm{B}_{\vec{r}}^{(\ell_{1},...,\ell_{d})}$ denotes a Pauli product on $R^{(\ell_{1},...,\ell_{d})}_{\vec{r}}$ as in $\bm{A}_{\vec{r}}^{(\ell_{1},...,\ell_{d})}$ of Eq.~\eqref{eq:Q_Expansion}.
The expansion coefficients $r_{\bm{B},\vec{r}}^{(\ell_{1},...,\ell_{d})}$ are given by linear combinations of coefficients $c_{\bm{A},\vec{r}}^{(\ell_{1},...,\ell_{d})}$. 
We divide the commutator by $2i$ to simplify the expression of $r_{\bm{B},\vec{r}}^{(\ell_{1},...,\ell_{d})}$.
In order for $Q$ to commute with $H$, they must satisfy
\begin{align}
    r_{\bm{B},\vec{r}}^{(\ell_{1},...,\ell_{d})}=0\quad\text{for all }\bm{B}_{\vec{r}}^{(\ell_{1},...,\ell_{d})},
    \label{eq:ConservCond}
\end{align}
where $\ell_{\mu}\le k_{\mu}+1$  ($\mu=1,...,d$).

In the following sections, we will show that, 
if one of $k_{1},...,k_{d}$ (say $k_{\mu^*}$) satisfies $3\le k_{\mu^*}\le L/2$, 
then all $c_{\bm{A},\vec{r}}^{(\ell_{1},...,\ell_{d})}$ with $\ell_{\mu^*}=k_{\mu^*}$ must be zero in order for Eq.~\eqref{eq:ConservCond} to be satisfied. 
This means that $k_{\mu^*}$ can be taken smaller, and indicates absence of $(k_{1},...,k_{d})$-local conserved quantities 
satisfying $3\le k_{\mu^*}\le L/2$.
In addition, in the case of $k_{\mu^*}\le 2$, we investigate the coefficients $c_{\bm{A},\vec{r}}^{(\ell_{1},...,\ell_{d})}$ 
and show that 
$(k_1,...,k_d)$-local conserved quantities
corresponding to the solutions of Eq.~\eqref{eq:ConservCond}
is restricted to a linear combination of the Hamiltonian~$H$ and the identity~$I$.

\section{\label{sec:2D}Proof in two dimension}

In this section, we consider the model~\eqref{eq:Hamiltonian} with $d=2$ (i.e., the mixed-field Ising model on the square lattice). 
In order to analyze Eq.~\eqref{eq:ConservCond}, we need to calculate commutators of two Pauli products, as in
\begin{align}
    \frac{1}{2i}[c_{\bm{A},\vec{r}}^{(1,2)}Y_{\vec{r}}X_{\vec{r}+\vec{e}_{2}},J^{1}Z_{\vec{r}}Z_{\vec{r}+\vec{e}_{1}}]&=J^{1}c_{\bm{A},\vec{r}}^{(1,2)}X_{\vec{r}}X_{\vec{r}+\vec{e}_{2}}Z_{\vec{r}+\vec{e}_{1}},
    \label{eq:DEF_diag_J1}\\
    \frac{1}{2i}[c_{\bm{A},\vec{r}}^{(1,2)}Y_{\vec{r}}X_{\vec{r}+\vec{e}_{2}},h^{x}X_{\vec{r}}]&=-h^{x}c_{\bm{A},\vec{r}}^{(1,2)}Z_{\vec{r}}X_{\vec{r}+\vec{e}_{2}},
    \label{eq:DEF_diag_hx}\\
    \frac{1}{2i}[c_{\bm{A},\vec{r}}^{(1,2)}Y_{\vec{r}}X_{\vec{r}+\vec{e}_{2}},h^{z}Z_{\vec{r}}]&=h^{z}c_{\bm{A},\vec{r}}^{(1,2)}X_{\vec{r}}X_{\vec{r}+\vec{e}_{2}},
    \label{eq:DEF_diag_hz}
\end{align}
We represent such calculations, 
Eqs.~\eqref{eq:DEF_diag_J1}, \eqref{eq:DEF_diag_hx}, and \eqref{eq:DEF_diag_hz}, 
diagrammatically as
\begin{align}
    \begin{array}{||c||c}\ucline{1-1}\hline
    Y_{\vec{r}}&\cI{X}\\\hline
     & \\\cline{1-1}\dcline{1-1}
    \end{array}
    \ &\to \
    \begin{array}{|cc|} \hline
    X_{\vec{r}}&X\\
    Z& \\\hline
    \end{array},
    \label{eq:DEF_diag_J1_2}\\[5pt]
    \begin{array}{|cc|}\hline
    \Igray{Y_{\vec{r}}}&X\\\hline
    \end{array}
    \ &\to \
    \begin{array}{|cc|} \hline
    Z_{\vec{r}}&X\\\hline
    \end{array},
    \label{eq:DEF_diag_hx_2}\\[5pt]
    \begin{array}{||c||c|}\ucline{1-1}\hline
    Y_{\vec{r}}&X\\\hline\dcline{1-1}
    \end{array}
    \ &\to \
    \begin{array}{|cc|} \hline
    X_{\vec{r}}&X\\\hline
    \end{array},
    \label{eq:DEF_diag_hz_2}
\end{align}
respectively.
Here $\vec{e}_{1}$ is chosen downward and $\vec{e}_{2}$ rightward.
The symbols (the Pauli operators) on the left-hand side (LHS) of the diagram represent the term coming from $Q$, and 
those on the right-hand side (RHS) represent the term contributing to $[Q, H]/2i$.
We call the former the \emph{input} and the latter the \emph{output}.
The terms coming from $H$, the $ZZ$, $X$, and $Z$ terms, are represented by 
the double-border rectangle, the gray-filled square, and the double-border square
on the LHS, respectively. 
The single-border rectangle on the LHS (resp. the RHS) represents the smallest rectangle 
that contains the support of 
the input (resp. the output).
We say that some input or output is an $(\ell_{1},\ell_{2})$-support one if 
this rectangle
is given by $R^{(\ell_{1},\ell_{2})}_{\vec{r}}$ for some $\vec{r}$. 
For instance, the Pauli product $X_{\vec{r}}X_{\vec{r}+\vec{e}_{2}}Z_{\vec{r}+\vec{e}_{1}}$ represented on the RHS of Eq.~\eqref{eq:DEF_diag_J1_2} is a $(2,2)$-support output.
As in Eq.~\eqref{eq:DEF_diag_J1_2}, we omit the identities $I$ in the diagram unless necessary. We also omit the site index of each Pauli operator unless necessary.

Now we introduce some terminology to classify Pauli products. 
Without loss of generality, we take  $\mu^*=1$, that is, we assume that $k_{1}\le L/2$ while $k_{2}$ is arbitrary.
Take an arbitrary Pauli product $\bm{A}^{(\ell_{1},\ell_{2})}_{\vec{r}}$.
The following two edges of the rectangle $R^{(\ell_{1},\ell_{2})}_{\vec{r}}$ play a crucial role in the proof,
\begin{align}
    E^{(\ell_{1},\ell_{2})}_{\bm{A},\vec{r}}&=\{(r_{1},p_{2})|r_{2}\le p_{2}\le r_{2}+\ell_{2}-1\}\\
    F^{(\ell_{1},\ell_{2})}_{\bm{A},\vec{r}}&=\{(r_{1}+\ell_{1}-1,p_{2})|r_{2}\le p_{2}\le r_{2}+\ell_{2}-1\},
    \label{eq:DEF_Edges}
\end{align}
where $\vec{r}=(r_{1},r_{2})$.
(Since this section focuses on two-dimensional systems, we call these sets of sites ``edges'' rather than ``faces.'')
These edges correspond to the upper and the lower edges in the diagram.
For any Pauli product $\bm{A}^{(\ell_{1},\ell_{2})}_{\vec{r}}$, there is at least one Pauli operator $A_{\vec{p}}\neq I_{\vec{p}}$ on each edge, as explained below Eq.~\eqref{eq:PauliProd_A}.

This concept is useful to classify inputs 
whose side length in the direction $\vec{e}_{1}$ takes the maximum value $k_{1}$.
Take an arbitrary $(k_{1},\ell_{2})$-support input $\bm{A}^{(k_{1},\ell_{2})}_{\vec{r}}$,
where $\ell_{2}$ represents an arbitrary integer in $\{1,...,L\}$ since $k_{2}$ has no restriction.
By focusing on its edges $E^{(k_{1},\ell_{2})}_{\bm{A},\vec{r}}$ and $F^{(k_{1},\ell_{2})}_{\bm{A},\vec{r}}$,
it can be classified into one of the following three types:
\begin{enumerate}[label={\roman*.},ref={\roman*},align=right,leftmargin=!]
    \item \label{enum:input_XX}An input 
    that includes $X$ or $Y$ on both edges
    \item \label{enum:input_XZ}An input 
    that includes $X$ or $Y$ on one edge but not on the other edge
    \item \label{enum:input_ZZ}An input 
    that does not include $X$ nor $Y$ on both edges
\end{enumerate}

Using these, we explain the proof structure. 
See also Table~\ref{tbl:ProofStructure}.
The proof analyzes Eq.~\eqref{eq:ConservCond}: the condition that a candidate of a $(k_{1},k_{2})$-local conserved quantity $Q$ given in Eq.~\eqref{eq:Q_Expansion} commutes with $H$. 
We divide the proof into four parts. 
The first, second, and third parts (Secs.~\ref{sec:2D_XX}, \ref{sec:2D_XZ}, and \ref{sec:2D_ZZ}) investigate $(k_{1},\ell_{2})$-support inputs of type~\ref{enum:input_XX}, \ref{enum:input_XZ}, and \ref{enum:input_ZZ}, respectively.
By showing the corresponding coefficients $c_{\bm{A},\vec{r}}^{(k_{1},\ell_{2})}$ to be zero,
these analyses prove absence of $(k_{1},k_{2})$-local conserved quantities with $3\le k_{1}\le L/2$. 
The fourth part (Sec.~\ref{sec:2D_k_le_2}) investigates the $k_{1}\le 2$ case in detail and proves that any $(k_{1},k_{2})$-local conserved quantities with $k_{1}\le 2$ are restricted to 
the Hamiltonian $H$.
Note that although this proof structure is very similar to the previous proof in one dimension~\cite{Chiba2024}, the result of this paper does not follow directly from the previous proof 
because we need to consider various inputs that have no counterparts in one dimension.

\begin{table}
\caption{\label{tbl:ProofStructure}%
Structure of the proof in two dimension.} 
\begin{ruledtabular}
\begin{tabular}{ll}
Section &  Inputs examined
\\ \hline
\ref{sec:2D_XX} & Type~\ref{enum:input_XX}: Includes $X$ or $Y$ on both edges
\\[3pt] 
\ref{sec:2D_XZ} & Type~\ref{enum:input_XZ}: Includes $X$ or $Y$ on one edge (non-$Z$ edge)
\\
 & \hspace{30pt} but not on the other edge ($Z$ edge)
\\[3pt] 
 & Type~\ref{enum:input_XZ_Zmulti}: Includes more than one $Z$ on its $Z$ edge
\\[3pt] 
 & Type~\ref{enum:input_XZ_Zone}: Includes exactly one $Z$ on its $Z$ edge
 \\[3pt] 
\ref{sec:2D_XZ_Z(X)Y} & Type-\ref{enum:input_XZ_Zone} inputs of the form $Z(X)^{k_{1}-2}Y$,\, $Y(X)^{k_{1}-2}Z$
\\[3pt]
\ref{sec:2D_ZZ} & Type~\ref{enum:input_ZZ}: Includes neither $X$ nor $Y$ on both edges
\\[3pt] 
 & Type~\ref{enum:input_ZZ_Zmulti}: Includes more than one $Z$ on either edge
\\[3pt] 
 & Type~\ref{enum:input_ZZ_Zone}: Includes exactly one $Z$ on both edges
 \\[3pt] 
\ref{sec:2D_ZZ_Z(X)Z} & Type-\ref{enum:input_ZZ_Zone} input of the form $Z(X)^{k_{1}-2}Z$
\\[3pt]
\ref{sec:2D_k_le_2} & All inputs in the case of $k_{1}\le 2$
\\
\end{tabular}
\end{ruledtabular}
\end{table}%

\subsection{\label{sec:2D_XX}Type~\ref{enum:input_XX}: Inputs that include $X$ or $Y$ on both edges}

This subsection investigates all $(k_{1},\ell_{2})$-support inputs $\bm{A}^{(k_{1},\ell_{2})}_{\vec{r}}$ of type~\ref{enum:input_XX} in the case of $3\le k_{1}\le L/2$,
and shows $c_{\bm{A},\vec{r}}^{(k_{1},\ell_{2})}=0$ for all such inputs. 

As an example, we consider the $(k_{1},\ell_{2})$-support input 
of the form
\begin{align}
    \begin{array}{|cccc|}\hline
    X_{\vec{r}}&\ast&\ast&\ast\\
    \ast&\ast&\ast&\ast\\
    \ast&\ast&X&\ast\\\hline
    \end{array}.
    \label{eq:Example_InputXX}
\end{align}
Here each asterisk~``$\ast$'' represents an arbitrary operator $X_{\vec{p}},Y_{\vec{p}},Z_{\vec{p}},I_{\vec{p}}$ on each site~$\vec{p}$ (or those on multiple sites).
Note that both upper and lower edges in this diagram can include some $Z$'s.

By applying the $ZZ$ term (coming from $H$) to a non-$Z$ site on the lower edge 
$F^{(k_{1},\ell_{2})}_{\bm{A},\vec{r}}$, we have the following diagram
\begin{align}
    \begin{array}{cccc}\hline
    \Ic{X_{\vec{r}}}&\ast&\ast&\cI{\ast}\\
    \Ic{\ast}&\ast&\ast&\cI{\ast}\\\ucline{3-3}\cline{3-3}
    \Ic{\ast}&\ast&\IIcII{X}&\cI{\ast}\\\hline
    &&\IIcII{}&\\\cline{3-3}\dcline{3-3}
    \end{array}
     \ &\to \ 
    \begin{array}{|cccc|}\hline
    X_{\vec{r}}&\ast&\ast&\ast\\
    \ast&\ast&\ast&\ast\\
    \ast&\ast&Y&\ast\\
    &&Z&\\\hline
    \end{array}.
    \label{eq:diag_InputXX}
\end{align}
Since the above output is a $(k_{1}+1,\ell_{2})$-support one, 
the contribution to this output is restricted to $(k_{1},\ell_{2})$-support inputs applied by the $ZZ$ term to their upper or lower edge.
However, 
there are no contributions from the upper-edge case because the upper edge of this output includes a non-$Z$ operator (such as $X_{\vec{r}}$), which cannot be produced by applying $ZZ$ term to the upper edge of any $(k_{1},\ell_{2})$-support input.
In addition, we can directly verify that there are no other contributions from the lower-edge case.
Thus, Eq.~\eqref{eq:ConservCond} reduces to
\begin{align}
    -J^{1} c_{\bm{A},\vec{r}}^{(k_{1},\ell_{2})}=0,
\end{align}
where the input $\bm{A}^{(k_{1},\ell_{2})}_{\vec{r}}$ is given by Eq.~\eqref{eq:Example_InputXX}.
The same argument can be applied to all other $(k_{1},\ell_{2})$-support inputs 
of type~\ref{enum:input_XX},
and we obtain the following proposition.
(Note that it is applicable even when $k_{1}=2$.)
\begin{proposition}[type-\ref{enum:input_XX} inputs]\label{prop:input_XX}
Assume $J^{1}\neq 0$. 
For $2\le k_{1}\le L/2$, any solution of Eq.~\eqref{eq:ConservCond} must satisfy
\begin{align}
    c_{\bm{A},\vec{r}}^{(k_{1},\ell_{2})}=0\quad\text{for any }\bm{A}^{(k_{1},\ell_{2})}_{\vec{r}}\text{ of type~\ref{enum:input_XX}}.
    \label{eq:proposition_InputXX}
\end{align}
Here the site $\vec{r}$ is arbitrary.
\end{proposition}
\noindent
Thus, the remaining $(k_{1},\ell_{2})$-support inputs are of type~\ref{enum:input_XZ} and \ref{enum:input_ZZ}.

\subsection{\label{sec:2D_XZ}Type~\ref{enum:input_XZ}: Inputs that include $X$ or $Y$ on one edge but not on the other edge}

This subsection investigates, in the case of $3\le k_{1}\le L/2$, all $(k_{1},\ell_{2})$-support inputs $\bm{A}^{(k_{1},\ell_{2})}_{\vec{r}}$ 
of type~\ref{enum:input_XZ}. 
Each input of type~\ref{enum:input_XZ} includes some $A_{\vec{p}}= X_{\vec{p}},Y_{\vec{p}}$ on one edge 
$E^{(k_{1},\ell_{2})}_{\bm{A},\vec{r}}$ or $F^{(k_{1},\ell_{2})}_{\bm{A},\vec{r}}$ 
but not on the other edge.
We call the former edge the ``non-$Z$ edge'' and the latter the ``$Z$ edge.''
Using this, inputs of type~\ref{enum:input_XZ} can be further divided into two types:
\begin{enumerate}[label={\ref{enum:input_XZ}-\alph*.},ref={\ref{enum:input_XZ}-\alph*},align=right,leftmargin=!]
    \item\label{enum:input_XZ_Zmulti}A type-\ref{enum:input_XZ} input that includes more than one $Z$ on its $Z$ edge
    \item\label{enum:input_XZ_Zone}A type-\ref{enum:input_XZ} input that includes exactly one $Z$ on its $Z$ edge
\end{enumerate}

For instance, 
\begin{align}
    \begin{array}{|cccc|}\hline
            Z_{\vec{r}}& &Z& \\
    \ast&\ast&\ast&\ast\\
    \ast&X&\ast&\ast\\\hline
    \end{array}
    \label{eq:Example_InputXZ_Zmulti}
\end{align}
is a type-\ref{enum:input_XZ_Zmulti} input,
and 
\begin{align}
        \begin{array}{|cccc|}\hline
    I_{\vec{r}}& &Z& \\
    \ast&\ast&\ast&\ast\\
    \ast&X&\ast&\ast\\\hline
    \end{array}
    \label{eq:Example_InputXZ_Zone}
\end{align}
is a type-\ref{enum:input_XZ_Zone} input.
Note that the lower edge in each diagram (the non-$Z$ edge) can include some $Z$'s.

First, we investigate inputs of type~\ref{enum:input_XZ_Zmulti}, taking the input~\eqref{eq:Example_InputXZ_Zmulti} as an example.
We apply the $ZZ$ term to one of the non-$Z$ sites on the non-$Z$ edge as in
\begin{align}
    \begin{array}{cccc}\hline
    \Ic{Z_{\vec{r}}}& &Z&\cI{}\\
    \Ic{\ast}&\ast&\ast&\cI{\ast}\\\ucline{2-2}
    \Ic{\ast}&\IIcII{X}&\ast&\cI{\ast}\\\hline
     &\IIcII{}& & \\\dcline{2-2}
    \end{array}
    \ &\to \
    \begin{array}{|cccc|}\hline
    Z_{\vec{r}}& &Z& \\
    \ast&\ast&\ast&\ast\\
    \ast&Y&\ast&\ast\\
     &Z& & \\\hline
    \end{array}.
\end{align}
We can see that there are no other inputs contributing to the above output because the upper edge includes multiple $Z$'s. 
Thus, Eq.~\eqref{eq:ConservCond} reduces to
\begin{align}
    -J^{1} c_{\bm{A},\vec{r}}^{(k_{1},\ell_{2})}=0,
\end{align}
where the input $\bm{A}^{(k_{1},\ell_{2})}_{\vec{r}}$ is given by Eq.~\eqref{eq:Example_InputXZ_Zmulti}.
The same argument can be applied to all other 
inputs of type~\ref{enum:input_XZ_Zmulti},
and we obtain the following proposition.
(Note that it is applicable even when $k_{1}=2$.)
\begin{proposition}[type-\ref{enum:input_XZ_Zmulti} inputs]\label{prop:input_XZ_Zmulti}
Assume $J^{1}\neq 0$. For $2\le k_{1}\le L/2$, any solution of Eq.~\eqref{eq:ConservCond} must satisfy
\begin{align}
    c_{\bm{A},\vec{r}}^{(k_{1},\ell_{2})}=0\quad\text{for any }\bm{A}^{(k_{1},\ell_{2})}_{\vec{r}}\text{ of type~\ref{enum:input_XZ_Zmulti}}.
\end{align}
\end{proposition}
\noindent
This proposition, combined with proposition~\ref{prop:input_XX} in Sec.~\ref{sec:2D_XX}, means that the remaining inputs are of type~\ref{enum:input_XZ_Zone} and of type~\ref{enum:input_ZZ}.

Next, we investigate 
type-\ref{enum:input_XZ_Zone} inputs.
For this purpose, 
we move on to the analysis of $(k_{1},\ell_{2})$-support outputs.
As an example, we consider the following diagram containing the input of the form~\eqref{eq:Example_InputXZ_Zone},
\begin{align}
        \begin{array}{cccc}\hline
    \Ic{I_{\vec{r}}}& &\gray{Z_{\vec{p}}}&\cI{}\\
    \Ic{\ast}&\ast&\ast&\cI{\ast}\\
    \Ic{\ast}&X_{\vec{q}}&\ast&\cI{\ast}\\\hline
    \end{array}
    \ &\to \
        \begin{array}{cccc}\hline
    \Ic{I_{\vec{r}}}& &Y_{\vec{p}}&\cI{}\\
    \Ic{\ast}&\ast&\ast&\cI{\ast}\\
    \Ic{\ast}&X_{\vec{q}}&\ast&\cI{\ast}\\\hline
    \end{array}.
    \label{eq:diag_inputXZ_Zone,X}
\end{align}
Importantly, the above output contains no contributions from $(k_{1}-1,\ell_{2})$-support inputs because both edges of this output include non-$Z$ operators.
Furthermore, this output does not contain contributions from $(k_{1},\ell_{2})$-support inputs of type~\ref{enum:input_ZZ} because, in the following diagram, no Hamiltonian terms can change both $A_{\vec{p}}=Z_{\vec{p}}, I_{\vec{p}}$ on the upper edge of the input and $A_{\vec{q}}=Z_{\vec{q}}, I_{\vec{q}}$ on the lower edge of the input to non-$Z$ operators simultaneously,
\begin{align}
    \begin{array}{|cccc|}\hline
    A_{\vec{r}}&\ast&A_{\vec{p}}&\ast\\
    \ast&\ast&\ast&\ast\\
    \ast&A_{\vec{q}}&\ast&\ast\\\hline
    \end{array}
    \ &\not\to \
    \begin{array}{|cccc|}\hline
    I_{\vec{r}}& &Y_{\vec{p}}& \\
    \ast&\ast&\ast&\ast\\
    \ast&X_{\vec{q}}&\ast&\ast\\\hline
    \end{array}.
\end{align}
In addition, the contributions from 
type-\ref{enum:input_XZ_Zone} inputs other than Eq.~\eqref{eq:diag_inputXZ_Zone,X} are 
absent because, in the following diagram, no Hamiltonian terms can change $A_{\vec{q}}=Z_{\vec{q}}, I_{\vec{q}}$ on the lower edge of the input to $X_{\vec{q}}$,
\begin{align}
    \begin{array}{|cccc|}\hline
    I_{\vec{r}}& &Y_{\vec{p}}& \\
    \ast&\ast&\ast&\ast\\
    \ast &A_{\vec{q}}&\ast&\ast\\\hline
    \end{array}
    \ \not\to \
    \begin{array}{|cccc|}\hline
    I_{\vec{r}}& &Y_{\vec{p}}& \\
    \ast&\ast&\ast&\ast\\
    \ast &X_{\vec{q}}&\ast&\ast\\\hline
    \end{array}.
    \label{eq:diag_inputXZ_Zone,X_2}
\end{align}
Therefore, Eq.~\eqref{eq:ConservCond} reduces to
\begin{align}
    h^{x} c_{\bm{A},\vec{r}}^{(k_{1},\ell_{2})}=0,
\end{align}
where the input $\bm{A}^{(k_{1},\ell_{2})}_{\vec{r}}$ is given by Eq.~\eqref{eq:Example_InputXZ_Zone}.
The same argument can be applied to all similar 
inputs of type~\ref{enum:input_XZ_Zone},
and we obtain the following lemma.
(Note that it is applicable even when $k_{1}=2$.)
\begin{lemma}\label{lemma:inputXZ_Zone_X}
Assume $J^{1},h^{x}\neq 0$ and $2\le k_{1}\le L/2$. 
Let $\bm{A}^{(k_{1},\ell_{2})}_{\vec{r}}$ be an arbitrary $(k_{1},\ell_{2})$-support input of type~\ref{enum:input_XZ_Zone}.
Any solution of Eq.~\eqref{eq:ConservCond} must satisfy
\begin{align}
    c_{\bm{A},\vec{r}}^{(k_{1},\ell_{2})}=0\quad &\text{for any type-\ref{enum:input_XZ_Zone} input }\bm{A}^{(k_{1},\ell_{2})}_{\vec{r}}\notag\\
    &\text{ that includes $X$ on its non-$Z$ edge}.
\end{align}
(For the definition of $Z$ and non-$Z$ edges, see the first paragraph of Sec.~\ref{sec:2D_XZ} .)
\end{lemma}
\noindent
This means that, for any remaining input of type~\ref{enum:input_XZ_Zone}, the non-$Z$ edge consists of only $Y_{\vec{q}}$, $Z_{\vec{q}}$, and $I_{\vec{q}}$.

\bigskip

Now we further investigate the remaining inputs of type~\ref{enum:input_XZ_Zone}, returning to the analysis of $(k_{1}+1,\ell_{2})$-support outputs.
As an example, we consider the following diagram,
\begin{align}
    \begin{array}{cccc}\hline
    \Ic{I_{\vec{r}}}& &Z_{\vec{p}}&\cI{} \\
    \Ic{\ast}&\ast& A_{\vec{p}+\vec{e}_{1}} &\cI{\ast}\\
    \Ic{\ast}&\ast&\ast&\cI{\ast}\\\ucline{2-2}
    \Ic{\ast}&\IIcII{Y}&\ast&\cI{\ast}\\\hline
    &\IIcII{}&&\\\dcline{2-2}
    \end{array}
    \ \to \
    \begin{array}{|cccc|}\hline
    I_{\vec{r}}& &Z_{\vec{p}}& \\
    \ast&\ast& A_{\vec{p}+\vec{e}_{1}} &\ast\\
    \ast&\ast&\ast&\ast\\
    \ast&X&\ast&\ast\\
    &Z&&\\\hline
    \end{array}.
    \label{eq:diag_InputXZ_Zone_Y,ZZ}
\end{align}

If $A_{\vec{p}+\vec{e}_{1}}=Z_{\vec{p}+\vec{e}_{1}},I_{\vec{p}+\vec{e}_{1}}$, the above output contains no other contributions.
In this case, Eq.~\eqref{eq:ConservCond} reduces to
\begin{align}
    J^{1} c_{\bm{A},\vec{r}}^{(k_{1},\ell_{2})}=0,
\end{align}
where the input $\bm{A}^{(k_{1},\ell_{2})}_{\vec{r}}$ is given by the LHS of Eq.~\eqref{eq:diag_InputXZ_Zone_Y,ZZ} with $A_{\vec{p}+\vec{e}_{1}}=Z_{\vec{p}+\vec{e}_{1}},I_{\vec{p}+\vec{e}_{1}}$.

If $A_{\vec{p}+\vec{e}_{1}}=Y_{\vec{p}+\vec{e}_{1}}$, the other contribution to the above output is 
\begin{align}
    \begin{array}{cccc}\ucline{3-3}
    I_{\vec{r}}& &\IIcII{}& \\\hline
    \Ic{A_{\vec{r}+\vec{e}_{1}}}&\ast& \IIcII{X_{\vec{p}+\vec{e}_{1}}} &\cI{\ast}\\\dcline{3-3}
    \Ic{\ast}&\ast&\ast&\cI{\ast}\\
    \Ic{\ast}&X&\ast&\cI{\ast}\\
    \Ic{}&Z&&\cI{}\\\hline
    \end{array}
    \ &\to \
    \begin{array}{|cccc|}\hline
    I_{\vec{r}}& &Z_{\vec{p}}& \\
    \ast&\ast& Y_{\vec{p}+\vec{e}_{1}} &\ast\\
    \ast&\ast&\ast&\ast\\
    \ast&X&\ast&\ast\\
    &Z&&\\\hline
    \end{array}.
    \label{eq:diag_InputXZ_X_Zone,ZZ}
\end{align}
In this case, Eq.~\eqref{eq:ConservCond} reduces to
\begin{align}
    J^{1} c_{\bm{A},\vec{r}}^{(k_{1},\ell_{2})}-J^{1} c_{\bm{B},\vec{r}+\vec{e}_{1}}^{(k_{1},\ell_{2})}=0,
\end{align}
where the input $\bm{A}^{(k_{1},\ell_{2})}_{\vec{r}}$ is given by the LHS of Eq.~\eqref{eq:diag_InputXZ_Zone_Y,ZZ} with $A_{\vec{p}+\vec{e}_{1}}=Y_{\vec{p}+\vec{e}_{1}}$ and the input $\bm{B}^{(k_{1},\ell_{2})}_{\vec{r}+\vec{e}_{1}}$ by the LHS of Eq.~\eqref{eq:diag_InputXZ_X_Zone,ZZ}.
However, 
the coefficient $c_{\bm{B},\vec{r}+\vec{e}_{1}}^{(k_{1},\ell_{2})}$
has been shown to be zero by Lemma~\ref{lemma:inputXZ_Zone_X}.
This means that $c_{\bm{A},\vec{r}}^{(k_{1},\ell_{2})}$ is also zero.
The same arguments can be applied to all other $(k_{1},\ell_{2})$-support inputs 
of type~\ref{enum:input_XZ_Zone},
and we obtain the following lemma.
\begin{lemma}\label{lemma:inputXZ_ZoneY_Y}
Assume $J^{1},h^{x}\neq 0$ and $3\le k_{1}\le L/2$. 
Let $\bm{A}^{(k_{1},\ell_{2})}_{\vec{r}}$ be an arbitrary $(k_{1},\ell_{2})$-support input of type~\ref{enum:input_XZ_Zone}, 
$\vec{p}$ be the single $Z$ site on the $Z$ edge of $\bm{A}^{(k_{1},\ell_{2})}_{\vec{r}}$ (i.e., $A_{\vec{p}}=Z_{\vec{p}}$),
and $\vec{q}$ be $\vec{q}=\vec{p}+\vec{e}_{1}$ if 
$\vec{p}\in E^{(k_{1},\ell_{2})}_{\bm{A},\vec{r}}$ 
while $\vec{q}=\vec{p}-\vec{e}_{1}$ if 
$\vec{p}\in F^{(k_{1},\ell_{2})}_{\bm{A},\vec{r}}$.
Any solution of Eq.~\eqref{eq:ConservCond} must satisfy
\begin{align}
    c_{\bm{A},\vec{r}}^{(k_{1},\ell_{2})}=0\quad \text{if }A_{\vec{q}}\neq X_{\vec{q}}.
\end{align}
\end{lemma}
\noindent
This lemma, combined with Lemma~\ref{lemma:inputXZ_Zone_X}, means that the remaining $(k_{1},\ell_{2})$-support inputs of type~\ref{enum:input_XZ_Zone} are of the form
\begin{align}
    \begin{array}{|cccc|}\hline
    I_{\vec{r}}& &Z_{\vec{p}}& \\
    \ast&\ast& X_{\vec{p}+\vec{e}_{1}} &\ast\\
    \ast&\ast&\ast&\ast\\
    \ast&Y&\ast&\ast\\\hline
    \end{array}
    \ \text{or} \
    \begin{array}{|cccc|}\hline
    A_{\vec{r}}&\ast&Y&\ast\\
    \ast&\ast&\ast&\ast\\
    \ast& X_{\vec{p}-\vec{e}_{1}} &\ast &\ast\\
     &Z_{\vec{p}}& & \\\hline
    \end{array},
\end{align}
where the non-$Z$ edges in both diagrams do not include any $X$'s.

For the output given in Eq.~\eqref{eq:diag_InputXZ_Zone_Y,ZZ}, if $A_{\vec{p}+\vec{e}_{1}}=X_{\vec{p}+\vec{e}_{1}}$, there is the other contribution,
\begin{align}
    \begin{array}{cccc}\ucline{3-3}
    I_{\vec{r}}& &\IIcII{} &\\\hline
    \Ic{A_{\vec{r}+\vec{e}_{1}}}&\ast& \IIcII{Y_{\vec{p}+\vec{e}_{1}}} &\cI{\ast}\\\dcline{3-3}
    \Ic{\ast}&\ast&\ast&\cI{\ast}\\
    \Ic{\ast}&X_{\vec{q}}&\ast&\cI{\ast}\\
    \Ic{}&Z_{\vec{q}+\vec{e}_{1}}&&\cI{}\\\hline
    \end{array}
    \ &\to \
    \begin{array}{|cccc|}\hline
    I_{\vec{r}}& &Z_{\vec{p}}& \\
    \ast&\ast& X_{\vec{p}+\vec{e}_{1}} &\ast\\
    \ast&\ast&\ast&\ast\\
    \ast&X_{\vec{q}}&\ast&\ast\\
    &Z_{\vec{q}+\vec{e}_{1}}&&\\\hline
    \end{array}.
    \label{eq:diag_InputXZ_Y_Zone,ZZ}
\end{align}
In this case, Eq.~\eqref{eq:ConservCond} reduces to
\begin{align}
    J^{1} c_{\bm{A},\vec{r}}^{(k_{1},\ell_{2})}+J^{1} c_{\bm{B},\vec{r}+\vec{e}_{1}}^{(k_{1},\ell_{2})}=0,
    \label{eq:c_InputXZ_ZoneX_Y,ZZ}
\end{align}
where the input $\bm{A}^{(k_{1},\ell_{2})}_{\vec{r}}$ is given by the LHS of Eq.~\eqref{eq:diag_InputXZ_Zone_Y,ZZ} with $A_{\vec{p}+\vec{e}_{1}}=X_{\vec{p}+\vec{e}_{1}}$ 
and the input $\bm{B}^{(k_{1},\ell_{2})}_{\vec{r}+\vec{e}_{1}}$ by the LHS of Eq.~\eqref{eq:diag_InputXZ_Y_Zone,ZZ}.

We also consider the following diagram where the remaining $(k_{1},\ell_{2})$-support input of type~\ref{enum:input_XZ_Zone} appears
\begin{align}
    \begin{array}{|cccc|}\hline
    I_{\vec{r}}& &\gray{Z_{\vec{p}}}& \\
    \ast&\ast& X_{\vec{p}+\vec{e}_{1}} &\ast\\
    \ast&\ast&\ast&\ast\\
    \ast&A_{\vec{q}-\vec{e}_{1}}&\ast&\ast\\
    \ast&Y_{\vec{q}}&\ast&\ast\\\hline
    \end{array}
    \ &\to \
    \begin{array}{|cccc|}\hline
    I_{\vec{r}}& &Y_{\vec{p}}& \\
    \ast&\ast& X_{\vec{p}+\vec{e}_{1}} &\ast\\
    \ast&\ast&\ast&\ast\\
    \ast&A_{\vec{q}-\vec{e}_{1}}&\ast&\ast\\
    \ast&Y_{\vec{q}}&\ast&\ast\\\hline
    \end{array}.
    \label{eq:diag_InputXZ_Zone_Y,X}
\end{align}
By almost the same reason explained below Eq.~\eqref{eq:diag_inputXZ_Zone,X}, the other contribution to the above output is given by
\begin{align}
    \begin{array}{|cccc|}\hline
    I_{\vec{r}}& &Y_{\vec{p}}& \\
    \ast&\ast& X_{\vec{p}+\vec{e}_{1}} &\ast\\
    \ast&\ast&\ast&\ast\\
    \ast&X_{\vec{q}-\vec{e}_{1}}&\ast&\ast\\
     &\gray{Z_{\vec{q}}}& & \\\hline
    \end{array}
    \ &\to \
    \begin{array}{|cccc|}\hline
    I_{\vec{r}}& &Y_{\vec{p}}& \\
    \ast&\ast& X_{\vec{p}+\vec{e}_{1}} &\ast\\
    \ast&\ast&\ast&\ast\\
    \ast&X_{\vec{q}-\vec{e}_{1}}&\ast&\ast\\
     &Y_{\vec{q}}& & \\\hline
    \end{array}.
    \label{eq:diag_InputXZ_Y_Zone,X}
\end{align}
Note that if the lower edge of the LHS of Eq.~\eqref{eq:diag_InputXZ_Zone_Y,X} includes some Pauli operators ($Y$ or $Z$) other than $Y_{\vec{q}}$, then the latter contribution~\eqref{eq:diag_InputXZ_Y_Zone,X} does not exist.
In addition, $A_{\vec{q}-\vec{e}_{1}}=X_{\vec{q}-\vec{e}_{1}}$ must hold in order for the above contribution to exist.
Therefore, Eq.~\eqref{eq:ConservCond} reduces to
\begin{align}
    h^{x} c_{\bm{A},\vec{r}}^{(k_{1},\ell_{2})}=
    \begin{cases}
        -h^{x}c_{\bm{B},\vec{r}}^{(k_{1},\ell_{2})} &\bigl(\text{if }A_{\vec{q}-\vec{e}_{1}}=X\text{ and }\\
        & A_{\vec{s}}=I\text{ for }\vec{s}\in F^{(k_{1},\ell_{2})}_{\bm{A},\vec{r}}\setminus\{\vec{q}\}\bigr)\\
        0 &(\text{otherwise})
    \end{cases},
    \label{eq:c_InputXZ_ZoneX_Y,X}
\end{align}
where the input $\bm{A}^{(k_{1},\ell_{2})}_{\vec{r}}$ is given by the LHS of Eq.~\eqref{eq:diag_InputXZ_Zone_Y,X}
and the input $\bm{B}^{(k_{1},\ell_{2})}_{\vec{r}}$ by the LHS of Eq.~\eqref{eq:diag_InputXZ_Y_Zone,X}.
Combining Eqs.~\eqref{eq:c_InputXZ_ZoneX_Y,ZZ} and \eqref{eq:c_InputXZ_ZoneX_Y,X}, we have 
\begin{align}
    c_{\bm{A},\vec{r}}^{(k_{1},\ell_{2})}=
    \begin{cases}
        -c_{\bm{C},\vec{r}-\vec{e}_{1}}^{(k_{1},\ell_{2})} &\bigl(\text{if }A_{\vec{q}-\vec{e}_{1}}=X\text{ and }\\
        & A_{\vec{s}}=I\text{ for }\vec{s}\in F^{(k_{1},\ell_{2})}_{\bm{A},\vec{r}}\setminus\{\vec{q}\}\bigr)\\
        0 &(\text{otherwise})
    \end{cases},
    \label{eq:ShiftRelation_InputXZ}
\end{align}
where the input $\bm{A}^{(k_{1},\ell_{2})}_{\vec{r}}$ is given by the LHS of Eq.~\eqref{eq:diag_InputXZ_Zone_Y,X}
and the input $\bm{C}^{(k_{1},\ell_{2})}_{\vec{r}-\vec{e}_{1}}$ is obtained from $\bm{A}^{(k_{1},\ell_{2})}_{\vec{r}}$ by adding $Z_{\vec{p}-\vec{e}_{1}}$, removing $Y_{\vec{q}}$, and replacing $Z_{\vec{p}}$ to $X_{\vec{p}}$ and $X_{\vec{q}-\vec{e}_{1}}$ to $Y_{\vec{q}-\vec{e}_{1}}$ (thus $\bm{C}^{(k_{1},\ell_{2})}_{\vec{r}-\vec{e}_{1}}$ is almost the one-site shift of $\bm{A}^{(k_{1},\ell_{2})}_{\vec{r}}$),
\begin{align}
    \bm{A}^{(k_{1},\ell_{2})}_{\vec{r}}=
    \begin{array}{|cccc|}\hline
    I_{\vec{r}}& &Z_{\vec{p}}& \\
    A_{\vec{r}+\vec{e}_{1}}&\ast& X_{\vec{p}+\vec{e}_{1}} &\ast\\
    \ast&\ast&\ast&\ast\\
    \ast&X_{\vec{q}-\vec{e}_{1}}&\ast&\ast\\
     &Y_{\vec{q}}& & \\\hline
    \end{array}
    \\
    \bm{C}^{(k_{1},\ell_{2})}_{\vec{r}-\vec{e}_{1}}=
    \begin{array}{|cccc|}\hline
    I_{\vec{r}-\vec{e}_{1}}& &Z_{\vec{p}-\vec{e}_{1}}& \\
     & & X_{\vec{p}} & \\
    A_{\vec{r}+\vec{e}_{1}}&\ast& X_{\vec{p}+\vec{e}_{1}} &\ast\\
    \ast&\ast&\ast&\ast\\
    \ast&Y_{\vec{q}-\vec{e}_{1}}&\ast&\ast\\\hline
    \end{array}.
\end{align}
Furthermore, by applying Eq.~\eqref{eq:ShiftRelation_InputXZ} to the coefficient $c_{\bm{C},\vec{r}-\vec{e}_{1}}^{(k_{1},\ell_{2})}$ instead of $c_{\bm{A},\vec{r}}^{(k_{1},\ell_{2})}$, we can show that $A_{\vec{q}-2\vec{e}_{1}}=X_{\vec{q}-2\vec{e}_{1}}$ must hold and that the lower edge of the input $\bm{C}^{(k_{1},\ell_{2})}_{\vec{r}-\vec{e}_{1}}$ must include only a single Pauli operator $Y_{\vec{q}-\vec{e}_{1}}$.
(Otherwise, both $c_{\bm{C},\vec{r}-\vec{e}_{1}}^{(k_{1},\ell_{2})}$ and $c_{\bm{A},\vec{r}}^{(k_{1},\ell_{2})}$ must be zero.)
By repeating such a discussion, we obtain the following proposition.
\begin{proposition}[type-\ref{enum:input_XZ_Zone} inputs]\label{prop:input_XZ_Zone}
Assume $J^{1},h^{x}\neq 0$ and $3\le k_{1}\le L/2$. 
Let $\bm{A}^{(k_{1},\ell_{2})}_{\vec{r}}$ be an arbitrary $(k_{1},\ell_{2})$-support input of type~\ref{enum:input_XZ_Zone}.
Any solution of Eq.~\eqref{eq:ConservCond} must satisfy $c_{\bm{A},\vec{r}}^{(k_{1},\ell_{2})}=0$ unless $\ell_{2}=1$ and
\begin{align}
    \bm{A}^{(k_{1},\ell_{2})}_{\vec{r}}=
    \begin{array}{|c|}\hline
    Z_{\vec{r}}\\
    (X)^{k_{1}-2}_{\vec{r}+\vec{e}_{1},\vec{e}_{1}}\\
    Y_{\vec{r}+(k_{1}-1)\vec{e}_{1}}\\\hline
    \end{array},\
    \begin{array}{|c|}\hline
    Y_{\vec{r}}\\
    (X)^{k_{1}-2}_{\vec{r}+\vec{e}_{1},\vec{e}_{1}}\\
    Z_{\vec{r}+(k_{1}-1)\vec{e}_{1}}\\\hline
    \end{array}.
    \label{eq:prop_InputXZ_Zone_1}
\end{align}
Here $(X)^{k_{1}-2}_{\vec{r}+\vec{e}_{1},\vec{e}_{1}}$ is defined by
\begin{align}
    (\bullet)^{k}_{\vec{r},\vec{e}}=\prod_{n=0}^{k-1}\bullet_{\vec{r}+n\vec{e}},
    \label{eq:DEF_(X)^k}
\end{align}
where $\bullet\in\{X,Y,Z,I\}$, $k=0,1,2,...$, and $\vec{r},\vec{e}\in\Lambda$.
In addition, the remaining coefficients of type~\ref{enum:input_XZ_Zone} are translation invariant in the direction~$\vec{e}_{1}$ and satisfy
\begin{align}
    c_{Z(X)^{k_{1}-2}Y,\vec{r}}^{(k_{1},1)}=-c_{Y(X)^{k_{1}-2}Z,\vec{r}}^{(k_{1},1)}=\text{const. indep. of }r_{1}.
    \label{eq:prop_InputXZ_Zone_2}
\end{align}
Here, $c_{Z(X)^{k_{1}-2}Y,\vec{r}}^{(k_{1},1)}$ and $c_{Y(X)^{k_{1}-2}Z,\vec{r}}^{(k_{1},1)}$ represent the coefficients corresponding to the first and the second inputs of the RHS of Eq.~\eqref{eq:prop_InputXZ_Zone_1}, respectively, 
and $r_{1}$ is the $1$st component of $\vec{r}$.
\end{proposition}
\noindent
This means that the remaining input of type~\ref{enum:input_XZ_Zone} is restricted to Eq.~\eqref{eq:prop_InputXZ_Zone_1}.
In addition, this proposition is consistent with the fact that the transverse Ising chain (the case of $d=1$ and $h^{z}=0$) has local conserved quantities of the form $\sum_{j}(Z_{j}X_{j+1}...X_{j+k-2}Y_{j+k-1}-Y_{j}X_{j+1}...X_{j+k-2}Z_{j+k-1})$~\cite{Chiba2024}
since the above analysis has not used any assumptions on the values of $J^{2}$ and $h^{z}$.

\bigskip

\subsubsection{\label{sec:2D_XZ_Z(X)Y}Inputs of the form \texorpdfstring{$Z(X)^{k_{1}-2}Y$ and $Y(X)^{k_{1}-2}Z$}{ZX...XY and YX...XZ}}

To conclude analysis of type-\ref{enum:input_XZ_Zone} inputs, this subsubsection shows that the coefficients $c_{Z(X)^{k_{1}-2}Y,\vec{r}}^{(k_{1},1)}$ and $c_{Y(X)^{k_{1}-2}Z,\vec{r}}^{(k_{1},1)}$ in Eq.~\eqref{eq:prop_InputXZ_Zone_2} must be zero, assuming $J^{2}\neq 0$.
(Note that the value of $h^{z}$ can be taken arbitrarily, including $h^{z}=0$.)
To this end, 
we analyze certain $(k_{1},\ell_{2})$-support outputs that contain only contributions from  the remaining $(k_{1},\ell_{2})$-support inputs~\eqref{eq:prop_InputXZ_Zone_1} of type~\ref{enum:input_XZ_Zone}, of type~\ref{enum:input_3D_ZZ}, and $(k_{1}-1,\ell_{2})$-support inputs.
The first output considered here is the following $(k_{1},2)$-support one
\begin{align}
    \begin{array}{|cc|}\hline
    Z_{\vec{r}}& \\
    (X)^{k_{1}-2}_{\vec{r}+\vec{e}_{1},\vec{e}_{1}}& \\
    X&Z\\\hline
    \end{array}.
    \label{eq:Output_Z(X)Y_1}
\end{align}
This output contains only the following contributions:
one from the remaining $(k_{1},1)$-support inputs of type~\ref{enum:input_XZ_Zone}, Eq.~\eqref{eq:prop_InputXZ_Zone_1},
\begin{align}
    \begin{array}{|c|c}\cline{1-1}
    Z_{\vec{r}}&\quad\\
    (X)^{k_{1}-2}_{\vec{r}+\vec{e}_{1},\vec{e}_{1}}& \\\ucline{1-2}
    \IIc{Y}&\IcII{}\\\dcline{1-2}
    \end{array},
\end{align}
and one from a $(k_{1}-1,2)$-support input 
[In the following of this subsubsection, this input is denoted by $(\bm{A}^1)^{(k_{1}-1,2)}_{\vec{r}+\vec{e}_{1}}$.]
\begin{align}
    \text{from }(\bm{A}^1)^{(k_{1}-1,2)}_{\vec{r}+\vec{e}_{1}}:\qquad 
    \begin{array}{cc}\ucline{1-1}
    \IIcII{}& \\\hline
    \IIcII{Y_{\vec{r}+\vec{e}_{1}}}&\cI{}\\\dcline{1-1}
    \Ic{(X)^{k_{1}-3}_{\vec{r}+2\vec{e}_{1},\vec{e}_{1}}}&\cI{}\\
    \Ic{X}&\cI{Z}\\\hline
    \end{array}.
\end{align}
[Note that contributions from $(k_{1},\ell_{2})$-support inputs of type~\ref{enum:input_ZZ} are absent because $A_{\vec{r}+(k_{1}-1)\vec{e}_{1}}=Z,I$ in such inputs cannot be changed to $X$.]
For this output, 
Eq.~\eqref{eq:ConservCond} reduces to
\begin{align}
    J^{2} c_{Z(X)^{k_{1}-2}Y,\vec{r}}^{(k_{1},1)}+J^{1} c_{(\bm{A}^1),\vec{r}+\vec{e}_{1}}^{(k_{1}-1,2)}=0.
    \label{eq:c_InputXZ_Z(X)Y_1}
\end{align}

The second output considered here is the following $(k_{1},2)$-support one
\begin{align}
    \begin{array}{|cc|}\hline
    Y_{\vec{r}+\vec{e}_{1}}& \\
    (X)^{k_{1}-3}_{\vec{r}+2\vec{e}_{1},\vec{e}_{1}}& \\
    Y&Z\\
    Z& \\\hline
    \end{array}.
    \label{eq:Output_Z(X)Y_2}
\end{align}
This output contains only the following contributions:
one from the remaining $(k_{1},1)$-support inputs of type~\ref{enum:input_XZ_Zone}, Eq.~\eqref{eq:prop_InputXZ_Zone_1},
\begin{align}
    \begin{array}{|c|c}\cline{1-1}
    Y_{\vec{r}+\vec{e}_{1}}&\quad\\
    (X)^{k_{1}-3}_{\vec{r}+2\vec{e}_{1},\vec{e}_{1}}& \\\ucline{1-2}
    \IIc{X}&\IcII{ }\\\dcline{1-2}
    Z& \\\cline{1-1}
    \end{array},
\end{align}
one from the $(k_{1}-1,2)$-support input $(\bm{A}^1)^{(k_{1}-1,2)}_{\vec{r}+\vec{e}_{1}}$
\begin{align}
    \text{from }(\bm{A}^1)^{(k_{1}-1,2)}_{\vec{r}+\vec{e}_{1}}:\qquad 
    \begin{array}{cc}\hline
    \Ic{Y_{\vec{r}+\vec{e}_{1}}}&\cI{}\\
    \Ic{(X)^{k_{1}-3}_{\vec{r}+2\vec{e}_{1},\vec{e}_{1}}}&\cI{}\\\ucline{1-1}
    \IIcII{X}&\cI{Z}\\\hline
    \IIcII{}& \\\dcline{1-1}
    \end{array},
\end{align}
and one from a $(k_{1},2)$-support input of type~\ref{enum:input_ZZ}
[In the following of this subsubsection, this input is denoted by $(\bm{B}^1)^{(k_{1},2)}_{\vec{r}+\vec{e}_{1}}$.]
\begin{align}
    \text{from }(\bm{B}^1)^{(k_{1},2)}_{\vec{r}+\vec{e}_{1}}:\qquad 
    \begin{array}{|cc|}\hline
    \Igray{Z_{\vec{r}+\vec{e}_{1}}}& \\
    (X)^{k_{1}-3}_{\vec{r}+2\vec{e}_{1},\vec{e}_{1}}& \\
    Y&Z\\
    Z& \\\hline
    \end{array}
\end{align}.
For this output, 
Eq.~\eqref{eq:ConservCond} reduces to
\begin{align}
    -J^{2} c_{Y(X)^{k_{1}-2}Z,\vec{r}+\vec{e}_{1}}^{(k_{1},1)}
    -J^{1} c_{(\bm{A}^1),\vec{r}+\vec{e}_{1}}^{(k_{1}-1,2)}
    +h^{x}c_{(\bm{B}^1),\vec{r}+\vec{e}_{1}}^{(k_{1},2)}
    =0.
    \label{eq:c_InputXZ_Z(X)Y_2}
\end{align}

The third output considered here is the following $(k_{1},2)$-support one
\begin{align}
    \begin{array}{|cc|}\hline
    Z_{\vec{r}+\vec{e}_{1}}& \\
    (X)^{k_{1}-3}_{\vec{r}+2\vec{e}_{1},\vec{e}_{1}}& \\
    Y&Z\\
    Y& \\\hline
    \end{array}.
    \label{eq:Output_Z(X)Y_3}
\end{align}
This output contains only the following contributions:
one from the remaining $(k_{1},1)$-support inputs of type~\ref{enum:input_XZ_Zone}, Eq.~\eqref{eq:prop_InputXZ_Zone_1},
\begin{align}
    \begin{array}{|c|c}\cline{1-1}
    Z_{\vec{r}+\vec{e}_{1}}&\quad\\
    (X)^{k_{1}-3}_{\vec{r}+2\vec{e}_{1},\vec{e}_{1}}& \\\ucline{1-2}
    \IIc{X}&\IcII{}\\\dcline{1-2}
    Y& \\\cline{1-1}
    \end{array},
\end{align}
one from a $(k_{1}-1,2)$-support input 
[In the following of this subsubsection, this input is denoted by $(\bm{A}^2)^{(k_{1}-1,2)}_{\vec{r}+2\vec{e}_{1}}$.]
\begin{align}
    \text{from }(\bm{A}^2)^{(k_{1}-1,2)}_{\vec{r}+2\vec{e}_{1}}:\qquad 
    \begin{array}{cc}\ucline{1-1}
    \IIcII{}& \\\hline
    \IIcII{Y_{\vec{r}+2\vec{e}_{1}}}&\cI{}\\\dcline{1-1}
    \Ic{(X)^{k_{1}-4}_{\vec{r}+3\vec{e}_{1},\vec{e}_{1}}}&\cI{}\\
    \Ic{Y}&\cI{Z}\\
    \Ic{Y}&\cI{}\\\hline
    \end{array},
\end{align}
and one from the $(k_{1},2)$-support input of type~\ref{enum:input_ZZ}, 
$(\bm{B}^1)^{(k_{1},2)}_{\vec{r}+\vec{e}_{1}}$,
\begin{align}
    \text{from }(\bm{B}^1)^{(k_{1},2)}_{\vec{r}+\vec{e}_{1}}:\qquad 
    \begin{array}{|cc|}\hline
    \Igray{Z_{\vec{r}+\vec{e}_{1}}}& \\
    (X)^{k_{1}-3}_{\vec{r}+2\vec{e}_{1},\vec{e}_{1}}& \\
    Y&Z\\
    Z& \\\hline
    \end{array}.
\end{align}
For this output, 
Eq.~\eqref{eq:ConservCond} reduces to
\begin{align}
    -J^{2} c_{Z(X)^{k_{1}-2}Y,\vec{r}+\vec{e}_{1}}^{(k_{1},1)}
    +J^{1} c_{(\bm{A}^2),\vec{r}+2\vec{e}_{1}}^{(k_{1}-1,2)}
    +h^{x}c_{(\bm{B}^1),\vec{r}+\vec{e}_{1}}^{(k_{1},2)}
    =0.
    \label{eq:c_InputXZ_Z(X)Y_3}
\end{align}
From Eqs.~\eqref{eq:c_InputXZ_Z(X)Y_1}, \eqref{eq:c_InputXZ_Z(X)Y_2}, \eqref{eq:c_InputXZ_Z(X)Y_3}, and \eqref{eq:prop_InputXZ_Zone_2}, we have
\begin{align}
    J^{1} c_{(\bm{A}^2),\vec{r}+2\vec{e}_{1}}^{(k_{1}-1,2)}
    =3J^{2} c_{Z(X)^{k_{1}-2}Y,\vec{r}}^{(k_{1},1)}.
    \label{eq:c_InputXZ_Z(X)Y_n=2}
\end{align}

Now let $n$ be an arbitrary integer in $\{2,3,...,k_{1}-2\}$.
The fourth output considered here is the following $(k_{1},2)$-support one
\begin{align}
    \begin{array}{|cc|}\hline
    Y_{\vec{r}+n\vec{e}_{1}}& \\
    (X)^{k_{1}-2-n}_{\vec{r}+(n+1)\vec{e}_{1},\vec{e}_{1}}& \\
    Y&Z\\
    (X)^{n-1}_{\vec{r}+k_{1}\vec{e}_{1},\vec{e}_{1}}& \\
    Z& \\\hline
    \end{array}.
    \label{eq:Output_Z(X)Y_4}
\end{align}
This output contains only the following contributions:
one from the remaining $(k_{1},1)$-support inputs of type~\ref{enum:input_XZ_Zone}, Eq.~\eqref{eq:prop_InputXZ_Zone_1},
\begin{align}
    \begin{array}{|c|c}\cline{1-1}
    Y_{\vec{r}+n\vec{e}_{1}}& \\
    (X)^{k_{1}-2-n}_{\vec{r}+(n+1)\vec{e}_{1},\vec{e}_{1}}& \\\ucline{1-2}
    \IIc{X}&\IcII{\quad }\\\dcline{1-2}
    (X)^{n-1}_{\vec{r}+k_{1}\vec{e}_{1},\vec{e}_{1}}& \\
    Z& \\\cline{1-1}
    \end{array},
\end{align}
one from a $(k_{1}-1,2)$-support input 
[In the following of this subsubsection, this input is denoted by $(\bm{A}^n)^{(k_{1}-1,2)}_{\vec{r}+n\vec{e}_{1}}$.]
\begin{align}
    \text{from }(\bm{A}^n)^{(k_{1}-1,2)}_{\vec{r}+n\vec{e}_{1}}:\qquad 
    \begin{array}{cc}\hline
    \Ic{Y_{\vec{r}+n\vec{e}_{1}}}&\cI{}\\
    \Ic{(X)^{k_{1}-2-n}_{\vec{r}+(n+1)\vec{e}_{1},\vec{e}_{1}}}&\cI{}\\
    \Ic{Y}&\cI{Z}\\
    \Ic{(X)^{n-2}_{\vec{r}+k_{1}\vec{e}_{1},\vec{e}_{1}}}&\cI{}\\\ucline{1-1}
    \IIcII{Y}&\cI{}\\\hline
    \IIcII{}& \\\dcline{1-1}
    \end{array},
\end{align}
and one from a $(k_{1},2)$-support input of type~\ref{enum:input_ZZ}
[In the following of this subsubsection, this input is denoted by $(\bm{B}^n)^{(k_{1},2)}_{\vec{r}+n\vec{e}_{1}}$.]
\begin{align}
    \text{from }(\bm{B}^n)^{(k_{1},2)}_{\vec{r}+n\vec{e}_{1}}:\qquad 
    \begin{array}{|cc|}\hline
    \Igray{Z_{\vec{r}+n\vec{e}_{1}}}& \\
    (X)^{k_{1}-2-n}_{\vec{r}+(n+1)\vec{e}_{1},\vec{e}_{1}}& \\
    Y&Z\\
    (X)^{n-1}_{\vec{r}+k_{1}\vec{e}_{1},\vec{e}_{1}}& \\
    Z& \\\hline
    \end{array}.
\end{align}
For this output, 
Eq.~\eqref{eq:ConservCond} reduces to
\begin{align}
    -J^{2} c_{Y(X)^{k_{1}-2}Z,\vec{r}+n\vec{e}_{1}}^{(k_{1},1)}
    +J^{1} c_{(\bm{A}^n),\vec{r}+n\vec{e}_{1}}^{(k_{1}-1,2)}& \notag\\
    +h^{x}c_{(\bm{B}^n),\vec{r}+n\vec{e}_{1}}^{(k_{1},2)}
    &=0,
    \label{eq:c_InputXZ_Z(X)Y_4}
\end{align}
where $n\in\{2,3,...,k_{1}-2\}$.

Next let $n$ be an arbitrary integer in $\{2,...,k_{1}-3\}$.
The fifth output considered here is the following $(k_{1},2)$-support one
\begin{align}
    \begin{array}{|cc|}\hline
    Z_{\vec{r}+n\vec{e}_{1}}& \\
    (X)^{k_{1}-2-n}_{\vec{r}+(n+1)\vec{e}_{1},\vec{e}_{1}}& \\
    Y&Z\\
    (X)^{n-1}_{\vec{r}+k_{1}\vec{e}_{1},\vec{e}_{1}}& \\
    Y& \\\hline
    \end{array}.
    \label{eq:Output_Z(X)Y_5}
\end{align}
This output contains only the following contributions:
one from the remaining $(k_{1},1)$-support inputs of type~\ref{enum:input_XZ_Zone}, Eq.~\eqref{eq:prop_InputXZ_Zone_1},
\begin{align}
    \begin{array}{|c|c}\cline{1-1}
    Z_{\vec{r}+n\vec{e}_{1}}& \\
    (X)^{k_{1}-2-n}_{\vec{r}+(n+1)\vec{e}_{1},\vec{e}_{1}}& \\\ucline{1-2}
    \IIc{X}&\IcII{\quad }\\\dcline{1-2}
    (X)^{n-1}_{\vec{r}+k_{1}\vec{e}_{1},\vec{e}_{1}}& \\
    Y& \\\cline{1-1}
    \end{array},
\end{align}
one from the $(k_{1}-1,2)$-support input 
$(\bm{A}^{n+1})^{(k_{1}-1,2)}_{\vec{r}+(n+1)\vec{e}_{1}}$
\begin{align}
    \text{from }(\bm{A}^{n+1})^{(k_{1}-1,2)}_{\vec{r}+(n+1)\vec{e}_{1}}:\qquad 
    \begin{array}{cc}\ucline{1-1}
    \IIcII{}& \\\hline
    \IIcII{Y_{\vec{r}+(n+1)\vec{e}_{1}}}&\cI{}\\\dcline{1-1}
    \Ic{(X)^{k_{1}-3-n}_{\vec{r}+(n+2)\vec{e}_{1},\vec{e}_{1}}}&\cI{}\\
    \Ic{Y}&\cI{Z}\\
    \Ic{(X)^{n-1}_{\vec{r}+k_{1}\vec{e}_{1},\vec{e}_{1}}}&\cI{} \\
    \Ic{Y}&\cI{}\\\hline
    \end{array},
\end{align}
and one from the $(k_{1},2)$-support input of type~\ref{enum:input_ZZ}, 
$(\bm{B}^n)^{(k_{1},2)}_{\vec{r}+n\vec{e}_{1}}$,
\begin{align}
    \text{from }(\bm{B}^n)^{(k_{1},2)}_{\vec{r}+n\vec{e}_{1}}:\qquad 
    \begin{array}{|cc|}\hline
    Z_{\vec{r}+n\vec{e}_{1}}& \\
    (X)^{k_{1}-2-n}_{\vec{r}+(n+1)\vec{e}_{1},\vec{e}_{1}}& \\
    Y&Z\\
    (X)^{n-1}_{\vec{r}+k_{1}\vec{e}_{1},\vec{e}_{1}}& \\
    \Igray{Z}& \\\hline
    \end{array}.
\end{align}
For this output, 
Eq.~\eqref{eq:ConservCond} reduces to
\begin{align}
    -J^{2} c_{Z(X)^{k_{1}-2}Y,\vec{r}+n\vec{e}_{1}}^{(k_{1},1)}
    +J^{1} c_{(\bm{A}^{n+1}),\vec{r}+(n+1)\vec{e}_{1}}^{(k_{1}-1,2)}& \notag\\
    +h^{x}c_{(\bm{B}^n),\vec{r}+n\vec{e}_{1}}^{(k_{1},2)}
    &=0,
    \label{eq:c_InputXZ_Z(X)Y_5}
\end{align}
where $n\in\{1,2,...,k_{1}-3\}$.
By almost the same calculation in the case of $n=k_{1}-2$, we can also show 
\begin{align}
    -J^{2} c_{Z(X)^{k_{1}-2}Y,\vec{r}+(k_{1}-2)\vec{e}_{1}}^{(k_{1},1)}
    -J^{1} c_{(\bm{A}^{k_{1}-1}),\vec{r}+(k_{1}-1)\vec{e}_{1}}^{(k_{1}-1,2)}& \notag\\
    +h^{x}c_{(\bm{B}^{k_{1}-2}),\vec{r}+(k_{1}-2)\vec{e}_{1}}^{(k_{1},2)}
    &=0,
    \label{eq:c_InputXZ_Z(X)Y_5_2}
\end{align}
where $(\bm{A}^{k_{1}-1})^{(k_{1}-1,2)}_{\vec{r}+(k_{1}-1)\vec{e}_{1}}$ denotes the following $(k_{1}-1,2)$-support input
\begin{align}
    (\bm{A}^{k_{1}-1})^{(k_{1}-1,2)}_{\vec{r}+(k_{1}-1)\vec{e}_{1}}\ =\ 
    \begin{array}{|cc|}\hline
    X_{\vec{r}+(k_{1}-1)\vec{e}_{1}}&Z\\
    (X)^{k_{1}-3}_{\vec{r}+k_{1}\vec{e}_{1},\vec{e}_{1}}& \\
    Y& \\\hline
    \end{array}.
\end{align}
From Eqs.~\eqref{eq:c_InputXZ_Z(X)Y_4}, \eqref{eq:c_InputXZ_Z(X)Y_5}, \eqref{eq:c_InputXZ_Z(X)Y_5_2}, and \eqref{eq:prop_InputXZ_Zone_2}, we have
\begin{align}
    J^{1} (c_{(\bm{A}^{n+1}),\vec{r}+(n+1)\vec{e}_{1}}^{(k_{1}-1,2)}-c_{(\bm{A}^{n}),\vec{r}+n\vec{e}_{1}}^{(k_{1}-1,2)})
    =2J^{2} c_{Z(X)^{k_{1}-2}Y,\vec{r}}^{(k_{1},1)},
    \label{eq:c_InputXZ_Z(X)Y_general_n}
\end{align}
for $n\in\{2,3,...,k_{1}-3\}$ and
\begin{align}
    &-J^{1} (c_{(\bm{A}^{k_{1}-1}),\vec{r}+(k_{1}-1)\vec{e}_{1}}^{(k_{1}-1,2)}
    +c_{(\bm{A}^{k_{1}-2}),\vec{r}+(k_{1}-2)\vec{e}_{1}}^{(k_{1}-1,2)})
    \notag\\
    &\qquad=2J^{2} c_{Z(X)^{k_{1}-2}Y,\vec{r}}^{(k_{1},1)},
    \label{eq:c_InputXZ_Z(X)Y_k1-2}
\end{align}
for $n=k_{1}-2$.

The sixth (the last) output considered here is the following $(k_{1},2)$-support one
\begin{align}
    \begin{array}{|cc|}\hline
    X_{\vec{r}+(k_{1}-1)\vec{e}_{1}}&Z\\
    (X)^{k_{1}-2}_{\vec{r}+k_{1}\vec{e}_{1},\vec{e}_{1}}& \\
    Z& \\\hline
    \end{array}.
    \label{eq:Output_Z(X)Y_6}
\end{align}
By a calculation similar to the above, we have
\begin{align}
    J^{2} c_{Y(X)^{k_{1}-2}Z,\vec{r}+(k_{1}-1)\vec{e}_{1}}^{(k_{1},1)}
    +J^{1} c_{(\bm{A}^{k_{1}-1}),\vec{r}+(k_{1}-1)\vec{e}_{1}}^{(k_{1}-1,2)}
    =0.
    \label{eq:c_InputXZ_Z(X)Y_6}
\end{align}

Using these results, we can show
\begin{align}
    0&=J^{1} c_{(\bm{A}^2),\vec{r}+2\vec{e}_{1}}^{(k_{1}-1,2)}\notag\\
    &\quad+J^{1}\sum_{n=2}^{k_{1}-3}(c_{(\bm{A}^{n+1}),\vec{r}+(n+1)\vec{e}_{1}}^{(k_{1}-1,2)}-c_{(\bm{A}^{n}),\vec{r}+n\vec{e}_{1}}^{(k_{1}-1,2)})\notag\\
    &\quad-J^{1} (c_{(\bm{A}^{k_{1}-1}),\vec{r}+(k_{1}-1)\vec{e}_{1}}^{(k_{1}-1,2)}
    +c_{(\bm{A}^{k_{1}-2}),\vec{r}+(k_{1}-2)\vec{e}_{1}}^{(k_{1}-1,2)})\notag\\
    &\quad+J^{1} c_{(\bm{A}^{k_{1}-1}),\vec{r}+(k_{1}-1)\vec{e}_{1}}^{(k_{1}-1,2)}\notag\\
    &=2(k_{1}-1)J^{2} c_{Z(X)^{k_{1}-2}Y,\vec{r}}^{(k_{1},1)}.
\end{align}
Here, the first equality is a trivial identity
and the second one follows from Eqs.~\eqref{eq:c_InputXZ_Z(X)Y_n=2}, \eqref{eq:c_InputXZ_Z(X)Y_general_n}, \eqref{eq:c_InputXZ_Z(X)Y_6}, and \eqref{eq:prop_InputXZ_Zone_2}.
(We can obtain the same equality even when $k_{1}=3,4$.)
Thus we obtain the following proposition.
\begin{proposition}[inputs $Z(X)^{k_{1}-2}Y$ and $Y(X)^{k_{1}-2}Z$]\label{prop:input_XZ_Z(X)Y}
Assume $J^{1},h^{x},J^{2}\neq 0$.
For $3\le k_{1}\le L/2$, 
any solution of Eq.~\eqref{eq:ConservCond} must satisfy 
\begin{align}
    c_{Z(X)^{k_{1}-2}Y,\vec{r}}^{(k_{1},1)}=c_{Y(X)^{k_{1}-2}Z,\vec{r}}^{(k_{1},1)}=0,
    \label{eq:prop_InputXZ_Z(X)Y}
\end{align}
where $c_{Z(X)^{k_{1}-2}Y,\vec{r}}^{(k_{1},1)}$ and $c_{Y(X)^{k_{1}-2}Z,\vec{r}}^{(k_{1},1)}$ represent the coefficients corresponding to the first and the second inputs of the RHS of Eq.~\eqref{eq:prop_InputXZ_Zone_1}, respectively.
\end{proposition}
\noindent
Thus, the remaining $(k_{1},\ell_{2})$-support inputs are restricted to of type~\ref{enum:input_ZZ}.
It is noteworthy that, in contrast to the one-dimensional case~\cite{Chiba2024}, we can show the absence of local conserved quantities of the form $\sum_{j}(Z_{j}X_{j+1}...X_{j+k-2}Y_{j+k-1}-Y_{j}X_{j+1}...X_{j+k-2}Z_{j+k-1})$ without assuming $h^{z}\neq 0$, thanks to the assumption $J^{2}\neq 0$.

\subsection{\label{sec:2D_ZZ}Type~\ref{enum:input_ZZ}: Inputs that do not include $X$ nor $Y$ on both edges}

This subsection investigates, in the case of $3\le k_{1}\le L/2$, all $(k_{1},\ell_{2})$-support inputs $\bm{A}^{(k_{1},\ell_{2})}_{\vec{r}}$ 
of type~\ref{enum:input_ZZ}. 
We can further divide such inputs into the following two types:
\begin{enumerate}[label={\ref{enum:input_ZZ}-\alph*.},ref={\ref{enum:input_ZZ}-\alph*},align=right,leftmargin=!]
    \item\label{enum:input_ZZ_Zmulti}A type-\ref{enum:input_ZZ} input that includes more than one $Z$ on either edge
    \item\label{enum:input_ZZ_Zone}A type-\ref{enum:input_ZZ} input that includes exactly one $Z$ on both edges
\end{enumerate}

For instance, 
\begin{align}
    \begin{array}{|cccc|}\hline
    Z_{\vec{r}}& &Z& \\
    \ast&\ast&\ast&\ast\\
     &Z& & \\\hline
    \end{array}
    \label{eq:Example_InputZZ_Zmulti}
\end{align}
is a type-\ref{enum:input_ZZ_Zmulti} input,
and 
\begin{align}
    \begin{array}{|cccc|}\hline
    I_{\vec{r}}& &Z& \\
    \ast&\ast&\ast&\ast\\
     &Z& & \\\hline
    \end{array}
    \label{eq:Example_InputZZ_Zone}
\end{align}
is a type-\ref{enum:input_ZZ_Zone} input.

First, we investigate inputs of type~\ref{enum:input_ZZ_Zmulti}, taking the input~\eqref{eq:Example_InputZZ_Zmulti} as an example.
We apply the $X$ term to 
a $Z$ site on the lower edge as in
\begin{align}
    \begin{array}{|cccc|}\hline
    Z_{\vec{r}}& &Z& \\
    \ast&\ast&\ast&\ast\\
     &\gray{Z}& & \\\hline
    \end{array}
    \ \to \
    \begin{array}{|cccc|}\hline
    Z_{\vec{r}}& &Z& \\
    \ast&\ast&\ast&\ast\\
     &Y& & \\\hline
    \end{array}.
    \label{eq:diag_InputZZ_Zmulti}
\end{align}
This output does not contain contributions from $(k_{1}-1,\ell_{2})$-support inputs because it includes multiple $Z$'s on the upper edge and a non-$Z$ operator on the lower edge.
Furthermore, contributions from $(k_{1},\ell_{2})$-support inputs of type~\ref{enum:input_ZZ} are restricted to only the above one because in order for any type-\ref{enum:input_ZZ} input to contribute to this output as in
\begin{align}
    \begin{array}{|cccc|}\hline
    A_{\vec{r}}&\ast&\ast&\ast\\
    \ast&\ast&\ast&\ast\\
    \ast&A_{\vec{p}}&\ast&\ast\\\hline
    \end{array}
    \ \to \
    \begin{array}{|cccc|}\hline
    Z_{\vec{r}}& &Z& \\
    \ast&\ast&\ast&\ast\\
     &Y_{\vec{p}}& & \\\hline
    \end{array},
\end{align}
(where $A_{\vec{p}}$ must be either $Z_{\vec{p}}$ or $I_{\vec{p}}$,) it is necessary to apply $X$ to the site $\vec{p}$, meaning that such an input must be exactly of the form~\eqref{eq:Example_InputZZ_Zmulti}.
Since the remaining $(k_{1},\ell_{2})$-support inputs are of type~\ref{enum:input_ZZ}, the above analysis completes all the contributions, and Eq.~\eqref{eq:ConservCond} reduces to
\begin{align}
    h^{x} c_{\bm{A},\vec{r}}^{(k_{1},\ell_{2})}=0,
\end{align}
where the input $\bm{A}^{(k_{1},\ell_{2})}_{\vec{r}}$ is given by Eq.~\eqref{eq:Example_InputZZ_Zmulti}.
The same argument can be applied to all other 
inputs of type~\ref{enum:input_ZZ_Zmulti},
and we obtain the following proposition.
\begin{proposition}[type-\ref{enum:input_ZZ_Zmulti} inputs]\label{prop:input_ZZ_Zmulti}
Assume $J^{1},h^{x},J^{2}\neq 0$. 
For $3\le k_{1}\le L/2$, any solution of Eq.~\eqref{eq:ConservCond} must satisfy
\begin{align}
    c_{\bm{A},\vec{r}}^{(k_{1},\ell_{2})}=0\quad\text{for any }\bm{A}^{(k_{1},\ell_{2})}_{\vec{r}}\text{ of type~\ref{enum:input_ZZ_Zmulti}}.
\end{align}
\end{proposition}
\noindent
Thus, the remaining $(k_{1},\ell_{2})$-support inputs are of type~\ref{enum:input_ZZ_Zone}.

Next, we investigate 
type-\ref{enum:input_ZZ_Zone} inputs,
taking the input~\eqref{eq:Example_InputZZ_Zone} as an example.
We consider the following diagram
\begin{align}
    \begin{array}{|cccc|}\hline
    I_{\vec{r}}& &Z_{\vec{p}}& \\
    \ast&\ast& A_{\vec{p}+\vec{e}_{1}} &\ast\\
    \ast&\ast&\ast&\ast\\
     &\gray{Z_{\vec{q}}}& & \\\hline
    \end{array}
    \ \to \
    \begin{array}{|cccc|}\hline
    I_{\vec{r}}& &Z_{\vec{p}}& \\
    \ast&\ast& A_{\vec{p}+\vec{e}_{1}} &\ast\\
    \ast&\ast&\ast&\ast\\
     &Y_{\vec{q}}& & \\\hline
    \end{array}.
    \label{eq:diag_inputZZ_Zone,X}
\end{align}
Importantly, the other $(k_{1},\ell_{2})$-support inputs of type~\ref{enum:input_ZZ_Zone} do not contribute to this output
because such a contribution takes the following form
\begin{align}
    \begin{array}{|cccc|}\hline
    A_{\vec{r}}&\ast&\ast&\ast\\
    \ast&\ast&\ast&\ast\\
    \ast&A_{\vec{q}}&\ast&\ast\\\hline
    \end{array}
    \ \to \
    \begin{array}{|cccc|}\hline
    I_{\vec{r}}& &Z& \\
    \ast&\ast&\ast&\ast\\
     &Y_{\vec{q}}& & \\\hline
    \end{array},
    \label{eq:diag_InputZZ_Zone_NoOther_ksupport}
\end{align}
(where $A_{\vec{q}}$ must be either $Z_{\vec{q}}$ or $I_{\vec{q}}$,) 
and hence it is necessary to apply $X$ to the site $\vec{q}$, 
meaning that such an input must be exactly of the form~\eqref{eq:Example_InputZZ_Zone}.
Furthermore, although $(k_{1}-1,\ell_{2})$-support inputs may contribute to the output of Eq.~\eqref{eq:diag_inputZZ_Zone,X}, it depends on $A_{\vec{p}+\vec{e}_{1}}$.

If $A_{\vec{p}+\vec{e}_{1}}=Z,I$, there are no contributions from $(k_{1}-1,\ell_{2})$-support inputs, resulting in 
\begin{align}
    h^{x} c_{\bm{A},\vec{r}}^{(k_{1},\ell_{2})}=0,
\end{align}
where the input $\bm{A}^{(k_{1},\ell_{2})}_{\vec{r}}$ is given by the LHS of Eq.~\eqref{eq:diag_inputZZ_Zone,X} with $A_{\vec{p}+\vec{e}_{1}}=Z,I$. 

If $A_{\vec{p}+\vec{e}_{1}}=Y$, there is the other contribution from a $(k_{1}-1,\ell_{2})$-support input
\begin{align}
    \begin{array}{cccc}\ucline{3-3}
     & &\IIcII{}& \\\hline
    \Ic{A_{\vec{r}+\vec{e}_{1}}}&\ast& \IIcII{X_{\vec{p}+\vec{e}_{1}}} &\cI{\ast}\\\dcline{3-3}
    \Ic{\ast}&\ast&\ast&\cI{\ast}\\
    \Ic{}&Y_{\vec{q}}& &\cI{}\\\hline
    \end{array}
    \ \to \
    \begin{array}{|cccc|}\hline
    I_{\vec{r}}& &Z_{\vec{p}}& \\
    \ast&\ast& Y_{\vec{p}+\vec{e}_{1}} &\ast\\
    \ast&\ast&\ast&\ast\\
     &Y_{\vec{q}}& & \\\hline
    \end{array}.
    \label{eq:diag_inputZZ_Zone,X_A=Y}
\end{align}
This means that Eq.~\eqref{eq:ConservCond} reduces to
\begin{align}
    h^{x} c_{\bm{A},\vec{r}}^{(k_{1},\ell_{2})}-J^{1}c_{\bm{B},\vec{r}+\vec{e}_{1}}^{(k_{1}-1,\ell_{2})}=0,
\end{align}
where the input $\bm{A}^{(k_{1},\ell_{2})}_{\vec{r}}$ is given by the LHS of Eq.~\eqref{eq:diag_inputZZ_Zone,X} with $A_{\vec{p}+\vec{e}_{1}}=Y$ 
and the input $\bm{B}^{(k_{1}-1,\ell_{2})}_{\vec{r}+\vec{e}_{1}}$ by the LHS of Eq.~\eqref{eq:diag_inputZZ_Zone,X_A=Y}. 
However, we can show $c_{\bm{B},\vec{r}+\vec{e}_{1}}^{(k_{1}-1,\ell_{2})}=0$ by considering the following diagram 
\begin{align}
    \begin{array}{cccc}\hline
    \Ic{A_{\vec{r}+\vec{e}_{1}}}&\ast& X_{\vec{p}+\vec{e}_{1}} &\cI{\ast}\\
    \Ic{\ast}&\ast&\ast&\cI{\ast}\\\ucline{2-2}
    \Ic{}&\IIcII{Y_{\vec{q}}}& &\cI{}\\\hline
      &\IIcII{} & & \\\dcline{2-2}
    \end{array}
    \ \to \
    \begin{array}{|cccc|}\hline
    A_{\vec{r}+\vec{e}_{1}}&\ast &X_{\vec{p}+\vec{e}_{1}}&\ast\\
    \ast&\ast& \ast &\ast\\
     &X_{\vec{q}}& & \\
     &Z_{\vec{q}+\vec{e}_{1}}& & \\\hline
    \end{array}.
\end{align}
Since 
the above output includes a non-$Z$ operator (namely $X_{\vec{p}+\vec{e}_{1}}$) on the upper edge, 
there are no other contributions from $(k_{1}-1,\ell_{2})$-support inputs.
In addition, any $(k_{1},\ell_{2})$-support inputs of type~\ref{enum:input_ZZ} cannot contribute to this output 
because in the following diagram no Hamiltonian terms can change $A_{\vec{p}+\vec{e}_{1}}=Z,I$ on the upper edge of the input 
to $X$,
\begin{align}
    \begin{array}{|cccc|}\hline
    A_{\vec{r}+\vec{e}_{1}}&\ast &A_{\vec{p}+\vec{e}_{1}}&\ast\\
    \ast&\ast& \ast &\ast\\
    \ast&\ast&\ast&\ast\\
    \ast&\ast&\ast&\ast\\\hline
    \end{array}
    \ \not\to \
    \begin{array}{|cccc|}\hline
    A_{\vec{r}+\vec{e}_{1}}&\ast &X_{\vec{p}+\vec{e}_{1}}&\ast\\
    \ast&\ast& \ast &\ast\\
     &X_{\vec{q}}& & \\
     &Z_{\vec{q}+\vec{e}_{1}}& & \\\hline
    \end{array}.
\end{align}
This means that, for this output, Eq.~\eqref{eq:ConservCond} reduces to
$J^{1}c_{\bm{B},\vec{r}+\vec{e}_{1}}^{(k_{1}-1,\ell_{2})}=0$,
resulting in
\begin{align}
    h^{x} c_{\bm{A},\vec{r}}^{(k_{1},\ell_{2})}=0,
\end{align}
where the input $\bm{A}^{(k_{1},\ell_{2})}_{\vec{r}}$ is given by the LHS of Eq.~\eqref{eq:diag_inputZZ_Zone,X} with $A_{\vec{p}+\vec{e}_{1}}=Y$.
The same argument can be applied to all similar 
inputs of type~\ref{enum:input_ZZ_Zone},
and we obtain the following lemma.
\begin{lemma}\label{lemma:input_ZZ_ZoneA_A=Y,Z,I}
Assume $J^{1},h^{x},J^{2}\neq 0$ and $3\le k_{1}\le L/2$. 
Let $\bm{A}^{(k_{1},\ell_{2})}_{\vec{r}}$ be an arbitrary $(k_{1},\ell_{2})$-support input of type~\ref{enum:input_ZZ_Zone}, 
and $\vec{p}$ 
(resp. $\vec{q}$) 
be the single $Z$ site on its edge 
$E^{(k_{1},\ell_{2})}_{\bm{A},\vec{r}}$ 
(resp. $F^{(k_{1},\ell_{2})}_{\bm{A},\vec{r}}$).
Any solution of Eq.~\eqref{eq:ConservCond} must satisfy
\begin{align}
    c_{\bm{A},\vec{r}}^{(k_{1},\ell_{2})}=0\quad \text{if }A_{\vec{p}+\vec{e}_{1}}\neq X\text{ or }A_{\vec{q}-\vec{e}_{1}}\neq X.
\end{align}
\end{lemma}
\noindent
This lemma means that the remaining inputs of type~\ref{enum:input_ZZ_Zone} 
satisfy both $A_{\vec{p}+\vec{e}_{1}}= X$ and $A_{\vec{q}-\vec{e}_{1}}= X$,
\begin{align}
    \begin{array}{|cccc|}\hline
    I_{\vec{r}}& &Z_{\vec{p}}& \\
    \ast&\ast& X_{\vec{p}+\vec{e}_{1}} &\ast\\
    \ast&\ast&\ast&\ast\\
    \ast& X_{\vec{q}-\vec{e}_{1}} &\ast &\ast\\
     &Z_{\vec{q}}& & \\\hline
    \end{array}.
    \label{eq:Example_InputZZ_Zone_ZX...XZ}
\end{align}

We further investigate the inputs of the form~\eqref{eq:Example_InputZZ_Zone_ZX...XZ}. 
We consider the following diagram
\begin{align}
    \begin{array}{|cccc|}\hline
    I_{\vec{r}}& &Z_{\vec{p}}& \\
    \ast&\ast& X_{\vec{p}+\vec{e}_{1}} &\ast\\
    \ast&\ast&\ast&\ast\\
    \ast& X_{\vec{q}-\vec{e}_{1}} &\ast &\ast\\
     &\gray{Z_{\vec{q}}}& & \\\hline
    \end{array}
    \ \to \ 
    \begin{array}{|cccc|}\hline
    I_{\vec{r}}& &Z_{\vec{p}}& \\
    \ast&\ast& X_{\vec{p}+\vec{e}_{1}} &\ast\\
    \ast&\ast&\ast&\ast\\
    \ast& X_{\vec{q}-\vec{e}_{1}} &\ast &\ast\\
     &Y_{\vec{q}}& & \\\hline
    \end{array}.
    \label{eq:diag_InputZZ_Zone_ZX...XZ_1}
\end{align}
This is the only contribution from $(k_{1},\ell_{2})$-support inputs of type~\ref{enum:input_ZZ_Zone}
because of the reason explained in Eq.~\eqref{eq:diag_InputZZ_Zone_NoOther_ksupport}.
In addition, the contributions from $(k_{1}-1,\ell_{2})$-support inputs are restricted to the following
\begin{align}
    \begin{array}{cccc}\ucline{3-3}
     & &\IIcII{}& \\\hline
    \Ic{A_{\vec{r}+\vec{e}_{1}}}&\ast& \IIcII{Y_{\vec{p}+\vec{e}_{1}}} &\cI{\ast}\\\dcline{3-3}
    \Ic{\ast}&\ast&\ast&\cI{\ast}\\
    \Ic{\ast}& X_{\vec{q}-\vec{e}_{1}} &\ast &\cI{\ast}\\
    \Ic{}&Y_{\vec{q}}& &\cI{} \\\hline
    \end{array}
    \ \to \ 
    \begin{array}{|cccc|}\hline
    I_{\vec{r}}& &Z_{\vec{p}}& \\
    \ast&\ast& X_{\vec{p}+\vec{e}_{1}} &\ast\\
    \ast&\ast&\ast&\ast\\
    \ast& X_{\vec{q}-\vec{e}_{1}} &\ast &\ast\\
     &Y_{\vec{q}}& & \\\hline
    \end{array}.
    \label{eq:diag_InputZZ_Zone_ZX...XZ_2}
\end{align}
Therefore, 
Eq.~\eqref{eq:ConservCond} for this output reduces to
\begin{align}
    h^{x} c_{\bm{A},\vec{r}}^{(k_{1},\ell_{2})}
    +J^{1}c_{\bm{B},\vec{r}+\vec{e}_{1}}^{(k_{1}-1,\ell_{2})}=0,
    \label{eq:c_InputZZ_Zone_ZX...XZ_1}
\end{align}
where the input $\bm{A}^{(k_{1},\ell_{2})}_{\vec{r}}$ is given by Eq.~\eqref{eq:Example_InputZZ_Zone_ZX...XZ} 
and the input $\bm{B}^{(k_{1}-1,\ell_{2})}_{\vec{r}+\vec{e}_{1}}$ by the LHS of Eq.~\eqref{eq:diag_InputZZ_Zone_ZX...XZ_2}. 
Furthermore, the input $\bm{B}^{(k_{1}-1,\ell_{2})}_{\vec{r}+\vec{e}_{1}}$ contributes to another $(k_{1},\ell_{2})$-support output
\begin{align}
    \begin{array}{cccc}\hline
    \Ic{A_{\vec{r}+\vec{e}_{1}}}&\ast& Y_{\vec{p}+\vec{e}_{1}} &\cI{\ast}\\
    \Ic{\ast}&\ast&\ast&\cI{\ast}\\
    \Ic{\ast}& X_{\vec{q}-\vec{e}_{1}} &\ast &\cI{\ast}\\\ucline{2-2}
    \Ic{}&\IIcII{Y_{\vec{q}}}& &\cI{}\\\hline
    &\IIcII{}& &\\\dcline{2-2}
    \end{array}
    \ \to \ 
    \begin{array}{|cccc|}\hline
    A_{\vec{r}+\vec{e}_{1}}&\ast& Y_{\vec{p}+\vec{e}_{1}} &\ast\\
    \ast&\ast&\ast&\ast\\
    \ast& X_{\vec{q}-\vec{e}_{1}} &\ast &\ast\\
     &X_{\vec{q}}& & \\
     &Z_{\vec{q}+\vec{e}_{1}}& & \\\hline
    \end{array}.
    \label{eq:diag_InputZZ_Zone_ZX...XZ_3}
\end{align}
If the upper edge of the above output consists of only a single Pauli operator $Y_{\vec{p}+\vec{e}_{1}}$ and the operator on the site $\vec{p}+2\vec{e}_{1}$ is $X$, then there is the other contribution from a type-\ref{enum:input_ZZ_Zone} input 
\begin{align}
    \begin{array}{|cccc|}\hline
    I_{\vec{r}+\vec{e}_{1}}& &\gray{Z_{\vec{p}+\vec{e}_{1}}}& \\
    \ast&\ast& X_{\vec{p}+2\vec{e}_{1}} &\ast\\
    \ast&\ast&\ast&\ast\\
    \ast& X_{\vec{q}-\vec{e}_{1}} &\ast &\ast\\
     &X_{\vec{q}}& & \\
     & Z_{\vec{q}+\vec{e}_{1}} & & \\\hline
    \end{array}
    \ \to \ 
    \begin{array}{|cccc|}\hline
    I_{\vec{r}+\vec{e}_{1}}& & Y_{\vec{p}+\vec{e}_{1}} & \\
    \ast&\ast& X_{\vec{p}+2\vec{e}_{1}} &\ast\\
    \ast&\ast&\ast&\ast\\
    \ast& X_{\vec{q}-\vec{e}_{1}} &\ast &\ast\\
     &X_{\vec{q}}& & \\
     &Z_{\vec{q}+\vec{e}_{1}}& & \\\hline
    \end{array}.
    \label{eq:diag_InputZZ_Zone_ZX...XZ_4}
\end{align}
In this case, 
Eq.~\eqref{eq:ConservCond} for this output reduces to
\begin{align}
    h^{x} c_{\bm{C},\vec{r}+\vec{e}_{1}}^{(k_{1},\ell_{2})}
    +J^{1}c_{\bm{B},\vec{r}+\vec{e}_{1}}^{(k_{1}-1,\ell_{2})}=0,
    \label{eq:c_InputZZ_Zone_ZX...XZ_2}
\end{align}
where the input $\bm{C}^{(k_{1},\ell_{2})}_{\vec{r}+\vec{e}_{1}}$ is given by the LHS of Eq.~\eqref{eq:diag_InputZZ_Zone_ZX...XZ_4}. 
On the other hand, if the upper edge of the output~\eqref{eq:diag_InputZZ_Zone_ZX...XZ_3} includes more than one Pauli operators or if the operator on the site $\vec{p}+2\vec{e}_{1}$ is not $X$, then there are no other contributions. This implies that $c_{\bm{B},\vec{r}+\vec{e}_{1}}^{(k_{1}-1,\ell_{2})}=c_{\bm{A},\vec{r}}^{(k_{1},\ell_{2})}=0$.
Combining these analyses, we have
\begin{align}
    c_{\bm{A},\vec{r}}^{(k_{1},\ell_{2})}=
    \begin{cases}
        c_{\bm{C},\vec{r}+\vec{e}_{1}}^{(k_{1},\ell_{2})} &\bigl(\text{if }A_{\vec{p}+2\vec{e}_{1}}=X\text{ and }\\
        & A_{\vec{s}+\vec{e}_{1}}=I\text{ for }\vec{s}\in 
        E^{(k_{1},\ell_{2})}_{\bm{A},\vec{r}}\setminus\{\vec{p}\}
        \bigr)\\
        0 &(\text{otherwise})
    \end{cases},
    \label{eq:ShiftRelation_InputZZ}
\end{align}
where the input $\bm{A}^{(k_{1},\ell_{2})}_{\vec{r}}$ is given by Eq.~\eqref{eq:Example_InputZZ_Zone_ZX...XZ}
and the input $\bm{C}^{(k_{1},\ell_{2})}_{\vec{r}+\vec{e}_{1}}$ by the LHS of Eq.~\eqref{eq:diag_InputZZ_Zone_ZX...XZ_4}.
Furthermore, by applying Eq.~\eqref{eq:ShiftRelation_InputZZ} to the coefficient $c_{\bm{C},\vec{r}+\vec{e}_{1}}^{(k_{1},\ell_{2})}$ instead of $c_{\bm{A},\vec{r}}^{(k_{1},\ell_{2})}$, 
we can show that $A_{\vec{p}+3\vec{e}_{1}}=X$ and 
$A_{\vec{s}+2\vec{e}_{1}}=I$ for all $\vec{s}\in E^{(k_{1},\ell_{2})}_{\bm{A},\vec{r}}\setminus\{\vec{p}\}$
must hold.
(Otherwise, both $c_{\bm{C},\vec{r}+\vec{e}_{1}}^{(k_{1},\ell_{2})}$ and $c_{\bm{A},\vec{r}}^{(k_{1},\ell_{2})}$ must be zero.)
By repeating such a discussion, we obtain the following proposition.
\begin{proposition}[type-\ref{enum:input_ZZ_Zone} inputs]\label{prop:input_ZZ_Zone}
Assume $J^{1},h^{x},J^{2}\neq 0$ and $3\le k_{1}\le L/2$. 
Let $\bm{A}^{(k_{1},\ell_{2})}_{\vec{r}}$ be an arbitrary $(k_{1},\ell_{2})$-support input of type~\ref{enum:input_ZZ_Zone}.
Any solution of Eq.~\eqref{eq:ConservCond} must satisfy $c_{\bm{A},\vec{r}}^{(k_{1},\ell_{2})}=0$ unless $\ell_{2}=1$ and
\begin{align}
    \bm{A}^{(k_{1},\ell_{2})}_{\vec{r}}=
    \begin{array}{|c|}\hline
    Z_{\vec{r}}\\
    (X)^{k_{1}-2}_{\vec{r}+\vec{e}_{1},\vec{e}_{1}}\\
    Z_{\vec{r}+(k_{1}-1)\vec{e}_{1}}\\\hline
    \end{array},
    \label{eq:prop_InputZZ_Zone_1}
\end{align}
where $(X)^{k_{1}-2}_{\vec{r}+\vec{e}_{1},\vec{e}_{1}}$ is defined by Eq.~\eqref{eq:DEF_(X)^k}.
In addition, the remaining coefficients of type~\ref{enum:input_ZZ_Zone} are translation invariant in the direction~$\vec{e}_{1}$,
\begin{align}
    c_{Z(X)^{k_{1}-2}Z,\vec{r}}^{(k_{1},1)}=\text{const. indep. of }r_{1}.
    \label{eq:prop_InputZZ_Zone_2}
\end{align}
Here, $c_{Z(X)^{k_{1}-2}Z,\vec{r}}^{(k_{1},1)}$ represents the coefficient corresponding to the input given by the RHS of Eq.~\eqref{eq:prop_InputZZ_Zone_1} and $r_{1}$ is the $1$st component of $\vec{r}$.
\end{proposition}
\noindent
This means that the remaining input of type~\ref{enum:input_ZZ_Zone} is restricted to Eq.~\eqref{eq:prop_InputZZ_Zone_1}.

\subsubsection{\label{sec:2D_ZZ_Z(X)Z}Inputs of the form \texorpdfstring{$Z(X)^{k_{1}-2}Z$}{ZX...XZ}}

To conclude analysis of $(k_{1},\ell_{2})$-support inputs of type~\ref{enum:input_ZZ_Zone}, 
we investigate the remaining type-\ref{enum:input_ZZ_Zone} input, Eq.~\eqref{eq:prop_InputZZ_Zone_1}, 
and show that the corresponding coefficient $c_{Z(X)^{k_{1}-2}Z,\vec{r}}^{(k_{1},1)}$ in Eq.~\eqref{eq:prop_InputZZ_Zone_2} must be zero.
From Eq.~\eqref{eq:c_InputZZ_Zone_ZX...XZ_1}, we have
\begin{align}
    h^{x}c_{Z(X)^{k_{1}-2}Z,\vec{r}}^{(k_{1},1)}+J^{1}c_{\bm{C},\vec{r}+\vec{e}_{1}}^{(k_{1}-1,1)}=0,
    \label{eq:c_Z(X)Z_Y(X)Y}
\end{align}
where the input $\bm{C}^{(k_{1}-1,1)}_{\vec{r}+\vec{e}_{1}}$ is given by
\begin{align}
    \bm{C}^{(k_{1}-1,1)}_{\vec{r}+\vec{e}_{1}}=
    \begin{array}{|c|}\hline
    Y_{\vec{r}+\vec{e}_{1}}\\
    (X)^{k_{1}-3}_{\vec{r}+2\vec{e}_{1},\vec{e}_{1}}\\
    Y_{\vec{r}+(k_{1}-1)\vec{e}_{1}}\\\hline
    \end{array}.
    \label{eq:Input_k-1_Y(X)Y}
\end{align}
In the following of this subsubsection, 
we analyze certain outputs that contain contributions from either the remaining type-\ref{enum:input_ZZ_Zone} input~\eqref{eq:prop_InputZZ_Zone_1} or the $(k_{1}-1,1)$-support input $\bm{C}^{(k_{1}-1,1)}_{\vec{r}+\vec{e}_{1}}$.

First we consider the following $(k_{1},2)$-support output
\begin{align}
    \begin{array}{|cc|}\hline
    Z_{\vec{r}}& \\
    (X)^{k_{1}-3}_{\vec{r}+\vec{e}_{1},\vec{e}_{1}}& \\
    Y&Z\\
    Z& \\\hline
    \end{array}.
\end{align}
This output contains the following contributions:
one from the $(k_{1},1)$-support input Eq.~\eqref{eq:prop_InputZZ_Zone_1},
\begin{align}
    \begin{array}{|c|c}\cline{1-1}
    Z_{\vec{r}}&\quad\\
    (X)^{k_{1}-3}_{\vec{r}+\vec{e}_{1},\vec{e}_{1}}& \\\ucline{1-2}
    \IIc{X}&\IcII{} \\\dcline{1-2}
    Z& \\\cline{1-1}
    \end{array},
\end{align}
one from a $(k_{1}-1,2)$-support input 
[In the following of this subsubsection, this input is denoted by $(\bm{A}^0)^{(k_{1}-1,2)}_{\vec{r}}$.]
\begin{align}
    \text{from }(\bm{A}^0)^{(k_{1}-1,2)}_{\vec{r}}:\qquad 
    \begin{array}{cc}\hline
    \Ic{Z_{\vec{r}}}&\cI{\quad}\\
    \Ic{(X)^{k_{1}-3}_{\vec{r}+\vec{e}_{1},\vec{e}_{1}}}&\cI{} \\\ucline{1-1}
    \IIcII{X}&\cI{Z}\\\hline
    \IIcII{}& \\\dcline{1-1}
    \end{array},
\end{align}
and one from a $(k_{1}-1,2)$-support input 
[In the following of this subsubsection, this input is denoted by $(\bm{B}^1)^{(k_{1}-1,2)}_{\vec{r}+\vec{e}_{1}}$.]
\begin{align}
    \text{from }(\bm{B}^1)^{(k_{1}-1,2)}_{\vec{r}+\vec{e}_{1}}:\qquad 
    \begin{array}{cc}\ucline{1-1}
    \IIcII{} & \\\hline
    \IIcII{Y_{\vec{r}+\vec{e}_{1}}}&\cI{}\\\dcline{1-1}
    \Ic{(X)^{k_{1}-4}_{\vec{r}+2\vec{e}_{1},\vec{e}_{1}}}&\cI{}\\
    \Ic{Y}&\cI{Z}\\
    \Ic{Z}&\cI{}\\\hline
    \end{array}.
\end{align}
For this output, 
Eq.~\eqref{eq:ConservCond} reduces to
\begin{align}
    -J^{2} c_{Z(X)^{k_{1}-2}Z,\vec{r}}^{(k_{1},1)}
    -J^{1} c_{(\bm{A}^0),\vec{r}}^{(k_{1}-1,2)}
    +J^{1} c_{(\bm{B}^1),\vec{r}+\vec{e}_{1}}^{(k_{1}-1,2)}
    =0.
    \label{eq:c_Z(X)Z_k_Output_1}
\end{align}

Now let $n$ be an arbitrary integer in $\{1,2,...,k_{1}-4\}$.
In a similar manner as above,
we consider the following $(k_{1},2)$-support output
\begin{align}
    \begin{array}{|cc|}\hline
    Z_{\vec{r}+n\vec{e}_{1}}& \\
    (X)^{k_{1}-3-n}_{\vec{r}+(n+1)\vec{e}_{1},\vec{e}_{1}}& \\
    Y&Z\\
    (X)^{n}_{\vec{r}+(k_{1}-1)\vec{e}_{1},\vec{e}_{1}}& \\
    Z& \\\hline
    \end{array}.
\end{align}
This output contains the following contributions:
one from the $(k_{1},1)$-support input Eq.~\eqref{eq:prop_InputZZ_Zone_1},
\begin{align}
    \begin{array}{|c|c}\cline{1-1}
    Z_{\vec{r}+n\vec{e}_{1}}& \\
    (X)^{k_{1}-3-n}_{\vec{r}+(n+1)\vec{e}_{1},\vec{e}_{1}}& \\\ucline{1-2}
    \IIc{X}&\IcII{\quad} \\\dcline{1-2}
    (X)^{n}_{\vec{r}+(k_{1}-1)\vec{e}_{1},\vec{e}_{1}}& \\
    Z& \\\cline{1-1}
    \end{array},
\end{align}
one from a $(k_{1}-1,2)$-support input 
[In the following of this subsubsection, this input is denoted by $(\bm{A}^n)^{(k_{1}-1,2)}_{\vec{r}+n\vec{e}_{1}}$.]
\begin{align}
    \text{from }(\bm{A}^n)^{(k_{1}-1,2)}_{\vec{r}+n\vec{e}_{1}}:\qquad 
    \begin{array}{cc}\hline
    \Ic{Z_{\vec{r}+n\vec{e}_{1}}}&\cI{} \\
    \Ic{(X)^{k_{1}-3-n}_{\vec{r}+(n+1)\vec{e}_{1},\vec{e}_{1}}}&\cI{}\\
    \Ic{Y}&\cI{Z}\\
    \Ic{(X)^{n-1}_{\vec{r}+(k_{1}-1)\vec{e}_{1},\vec{e}_{1}}}& \cI{}\\\ucline{1-1}
    \IIcII{Y}&\cI{} \\\hline
    \IIcII{}& \\\dcline{1-1}
    \end{array},
\end{align}
and one from a $(k_{1}-1,2)$-support input
[In the following of this subsubsection, this input is denoted by $(\bm{B}^{n+1})^{(k_{1}-1,2)}_{\vec{r}+(n+1)\vec{e}_{1}}$.]
\begin{align}
    \text{from }(\bm{B}^{n+1})^{(k_{1}-1,2)}_{\vec{r}+(n+1)\vec{e}_{1}}:\qquad 
    \begin{array}{cc}\ucline{1-1}
    \IIcII{}& \\\hline
    \IIcII{Y_{\vec{r}+(n+1)\vec{e}_{1}}}&\cI{} \\\dcline{1-1}
    \Ic{(X)^{k_{1}-4-n}_{\vec{r}+(n+2)\vec{e}_{1},\vec{e}_{1}}}& \cI{}\\
    \Ic{Y}&\cI{Z}\\
    \Ic{(X)^{n}_{\vec{r}+(k_{1}-1)\vec{e}_{1},\vec{e}_{1}}}& \cI{}\\
    \Ic{Z}& \cI{}\\\hline
    \end{array}.
\end{align}
For this output, 
Eq.~\eqref{eq:ConservCond} reduces to
\begin{align}
    -J^{2} c_{Z(X)^{k_{1}-2}Z,\vec{r}+n\vec{e}_{1}}^{(k_{1},1)}
    +J^{1} c_{(\bm{A}^n),\vec{r}+n\vec{e}_{1}}^{(k_{1}-1,2)}& \notag\\
    +J^{1}c_{(\bm{B}^{n+1}),\vec{r}+(n+1)\vec{e}_{1}}^{(k_{1}-1,2)}
    &=0,
    \label{eq:c_Z(X)Z_k_Output_2}
\end{align}
where $n\in\{1,2,...,k_{1}-4\}$.
By a similar calculation, we also have
\begin{align}
    -J^{2} c_{Z(X)^{k_{1}-2}Z,\vec{r}+(k_{1}-3)\vec{e}_{1}}^{(k_{1},1)}
    +J^{1} c_{(\bm{A}^{k_{1}-3}),\vec{r}+(k_{1}-3)\vec{e}_{1}}^{(k_{1}-1,2)}& \notag\\
    -J^{1}c_{(\bm{B}^{k_{1}-2}),\vec{r}+(k_{1}-2)\vec{e}_{1}}^{(k_{1}-1,2)}
    &=0,
    \label{eq:c_Z(X)Z_k_Output_3}
\end{align}
where the $(k_{1}-1,2)$-support inputs $(\bm{A}^{k_{1}-3})^{(k_{1}-1,2)}_{\vec{r}+(k_{1}-3)\vec{e}_{1}}$ and $(\bm{B}^{k_{1}-2})^{(k_{1}-1,2)}_{\vec{r}+(k_{1}-2)\vec{e}_{1}}$ are defined by
\begin{align}
    (\bm{A}^{k_{1}-3})^{(k_{1}-1,2)}_{\vec{r}+(k_{1}-3)\vec{e}_{1}} \ = \ 
    \begin{array}{|cc|}\hline
    Z_{\vec{r}+(k_{1}-3)\vec{e}_{1}}& \\
    Y&Z\\
    (X)^{k_{1}-4}_{\vec{r}+(k_{1}-1)\vec{e}_{1},\vec{e}_{1}}& \\
    Y& \\\hline
    \end{array},
\end{align}
and
\begin{align}
    (\bm{B}^{k_{1}-2})^{(k_{1}-1,2)}_{\vec{r}+(k_{1}-2)\vec{e}_{1}} \ = \ 
    \begin{array}{|cc|}\hline
    X_{\vec{r}+(k_{1}-2)\vec{e}_{1}}&Z\\
    (X)^{k_{1}-3}_{\vec{r}+(k_{1}-1)\vec{e}_{1},\vec{e}_{1}}& \\
    Z& \\\hline
    \end{array}.
\end{align}

\bigskip

We have investigated outputs that contain contributions from the input~\eqref{eq:prop_InputZZ_Zone_1} and obtained Eqs.~\eqref{eq:c_Z(X)Z_Y(X)Y}, \eqref{eq:c_Z(X)Z_k_Output_1}, \eqref{eq:c_Z(X)Z_k_Output_2}, and \eqref{eq:c_Z(X)Z_k_Output_3}.
To eliminate the coefficients of $(k_{1}-1,2)$-support inputs appearing in these equations, 
we will further analyze $(k_{1}-1,\ell_{2})$-support inputs in the following.

As in $(k_{1},\ell_{2})$-support inputs, 
we can classify $(k_{1}-1,\ell_{2})$-support inputs 
into three types: type~\ref{enum:input_XX}, \ref{enum:input_XZ} and \ref{enum:input_ZZ}.
For the following analysis, it is sufficient to examine only of type~\ref{enum:input_XX}.
As an example of $(k_{1}-1,\ell_{2})$-support inputs of type~\ref{enum:input_XX}, we consider the following form of inputs
\begin{align}
    \begin{array}{|cccc|}\hline
    Y_{\vec{r}}&\ast&\ast&\ast\\
    \ast&\ast&\ast&\ast\\
    \ast&\ast&Y&\ast\\\hline
    \end{array}.
    \label{eq:Example_k-1_InputXX}
\end{align}
This input contributes to a $(k_{1},\ell_{2})$-support output,
\begin{align}
    \begin{array}{cccc}\hline
    \Ic{Y_{\vec{r}}}&\ast&\ast&\cI{\ast}\\
    \Ic{\ast}&\ast&\ast&\cI{\ast}\\\ucline{3-3}
    \Ic{\ast}&\ast&\IIcII{Y}&\cI{\ast}\\\hline
    &&\IIcII{}&\\\dcline{3-3}
    \end{array}
    \ \to \ 
    \begin{array}{|cccc|}\hline
    Y_{\vec{r}}&\ast&\ast&\ast\\
    \ast&\ast&\ast&\ast\\
    \ast&\ast&X&\ast\\
    &&Z&\\\hline
    \end{array}.
    \label{eq:diag_k-1_InputXX}
\end{align}
Since the upper edge of this output includes a non-$Z$ operator (i.e., $Y_{\vec{r}}$), there are no other contributions from $(k_{1}-1,\ell_{2})$-support inputs.
Furthermore, the contributions from $(k_{1},\ell_{2})$-support inputs are restricted to the following form,
\begin{align}
    \begin{array}{|c|}\hline
    \IgrayI{Z_{\vec{r}}}\\
    (X)^{k_{1}-2}_{\vec{r}+\vec{e}_{1},\vec{e}_{1}}\\
    Z_{\vec{r}+(k_{1}-1)\vec{e}_{1}}\\\hline
    \end{array}
    \ \to \
    \begin{array}{|c|}\hline
    Y_{\vec{r}}\\
    (X)^{k_{1}-2}_{\vec{r}+\vec{e}_{1},\vec{e}_{1}}\\
    Z_{\vec{r}+(k_{1}-1)\vec{e}_{1}}\\\hline
    \end{array}.
\end{align}
In such a special case, Eq.~\eqref{eq:ConservCond} reduces to Eq.~\eqref{eq:c_Z(X)Z_Y(X)Y}.
Otherwise, Eq.~\eqref{eq:ConservCond} for the output~\eqref{eq:diag_k-1_InputXX} reduces to
\begin{align}
    J^{1} c_{\bm{A},\vec{r}}^{(k_{1}-1,\ell_{2})}=0,
\end{align}
where the input $\bm{A}^{(k_{1},\ell_{2})}_{\vec{r}}$ is given by Eq.~\eqref{eq:Example_k-1_InputXX}.
The same argument can be applied to all other $(k_{1}-1,\ell_{2})$-support inputs of type~\ref{enum:input_XX},
and we obtain the following lemma.
\begin{lemma}[$(k_{1}-1,\ell_{2})$-support inputs of type-\ref{enum:input_XX}]\label{lemma:k-1_input_XX}
Assume $J^{1},h^{x},J^{2}\neq 0$. 
For $3\le k_{1}\le L/2$, any solution of Eq.~\eqref{eq:ConservCond} must satisfy
\begin{align}
    c_{\bm{A},\vec{r}}^{(k_{1}-1,\ell_{2})}=0\quad\text{for any type-\ref{enum:input_XX} input }\bm{A}^{(k_{1}-1,\ell_{2})}_{\vec{r}}\neq \bm{C}^{(k_{1}-1,1)}_{\vec{r}},
    \label{eq:proposition_k-1_InputXX}
\end{align}
where the input $\bm{C}^{(k_{1}-1,1)}_{\vec{r}}$ is given by Eq.~\eqref{eq:Input_k-1_Y(X)Y}.
\end{lemma}
\noindent
This means that the remaining $(k_{1}-1,\ell_{2})$-support inputs of type-\ref{enum:input_XX} are 
only $\bm{C}^{(k_{1}-1,1)}_{\vec{r}}$.

Next we turn to the analysis of $(k_{1}-1,\ell_{2})$-support outputs that contain contributions from the input $\bm{C}^{(k_{1}-1,1)}_{\vec{r}}$.
We consider the following $(k_{1}-1,2)$-support output
\begin{align}
    \begin{array}{|cc|}\hline
    Y_{\vec{r}}& \\
    (X)^{k_{1}-3}_{\vec{r}+\vec{e}_{1},\vec{e}_{1}}& \\
    X&Z\\\hline
    \end{array}.
    \label{eq:Output_Z(X)Z_k-1_n=0}
\end{align}
Since this output includes non-$Z$ operators ($Y_{\vec{r}}$ and $X_{\vec{r}+(k_{1}-2)\vec{e}_{1}}$) on both edges, 
it contains no contributions from $(\ell_{1},\ell_{2})$-support inputs with $\ell_{1}\le k_{1}-2$.
In addition, it also contains no contribution from the remaining $(k_{1},\ell_{2})$-support input, Eq.~\eqref{eq:prop_InputZZ_Zone_1}.
These mean that the contributions may come only from $(k_{1}-1,\ell_{2})$-support inputs.
Furthermore, 
from lemma~\ref{lemma:k-1_input_XX}, 
any $(k_{1}-1,\ell_{2})$-support inputs other than the input $\bm{C}^{(k_{1}-1,1)}_{\vec{r}}$ 
are either of type~\ref{enum:input_XZ} or of type~\ref{enum:input_ZZ}, and hence they satisfy either
$A_{\vec{r}}=Z,I$ or $A_{\vec{r}+(k_{1}-2)\vec{e}_{1}}=Z,I$.
Thus, this output contains only the following contributions:
one from the $(k_{1}-1,1)$-support input $\bm{C}^{(k_{1}-1,1)}_{\vec{r}}$
\begin{align}
    \text{from }\bm{C}^{(k_{1}-1,1)}_{\vec{r}}:\qquad 
    \begin{array}{|c|c}\cline{1-1}
    Y_{\vec{r}}&\quad\\
    (X)^{k_{1}-3}_{\vec{r}+\vec{e}_{1},\vec{e}_{1}}& \\\ucline{1-2}
    \IIc{Y}&\IcII{} \\\dcline{1-2}
    \end{array},
\end{align}
and one from the $(k_{1}-1,2)$-support input 
$(\bm{A}^0)^{(k_{1}-1,2)}_{\vec{r}}$
\begin{align}
    \text{from }(\bm{A}^0)^{(k_{1}-1,2)}_{\vec{r}}:\qquad 
    \begin{array}{|cc|}\hline
    \Igray{Z_{\vec{r}}}&\quad\\
    (X)^{k_{1}-3}_{\vec{r}+\vec{e}_{1},\vec{e}_{1}}& \\
    X&Z\\\hline
    \end{array}.
\end{align}
For this output, 
Eq.~\eqref{eq:ConservCond} reduces to
\begin{align}
    J^{2} c_{\bm{C},\vec{r}}^{(k_{1}-1,2)}
    +h^{x} c_{(\bm{A}^0),\vec{r}}^{(k_{1}-1,2)}
    =0.
    \label{eq:c_Z(X)Z_k-1_Output_1}
\end{align}

Now let $n$ be an arbitrary integer in $\{1,2,...,k_{1}-3\}$.
In a similar manner as above,
we consider the following $(k_{1}-1,2)$-support output
\begin{align}
    \begin{array}{|cc|}\hline
    Y_{\vec{r}+n\vec{e}_{1}}& \\
    (X)^{k_{1}-3-n}_{\vec{r}+(n+1)\vec{e}_{1},\vec{e}_{1}}& \\
    Y&Z\\
    (X)^{n-1}_{\vec{r}+(k_{1}-1)\vec{e}_{1},\vec{e}_{1}}& \\
    Y& \\\hline
    \end{array}.
\end{align}
By the same reasons explained below Eq.~\eqref{eq:Output_Z(X)Z_k-1_n=0}, this output contains only the following contributions:
one from the $(k_{1}-1,1)$-support input $\bm{C}^{(k_{1}-1,1)}_{\vec{r}+n\vec{e}_{1}}$
\begin{align}
    \text{from }\bm{C}^{(k_{1}-1,1)}_{\vec{r}+n\vec{e}_{1}}:\qquad 
    \begin{array}{|c|c}\cline{1-1}
    Y_{\vec{r}+n\vec{e}_{1}}& \\
    (X)^{k_{1}-3-n}_{\vec{r}+(n+1)\vec{e}_{1},\vec{e}_{1}}& \\\ucline{1-2}
    \IIc{X}&\IcII{\quad} \\\dcline{1-2}
    (X)^{n-1}_{\vec{r}+(k_{1}-1)\vec{e}_{1},\vec{e}_{1}}& \\
    Y& \\\cline{1-1}
    \end{array},
\end{align}
one from the $(k_{1}-1,2)$-support input 
$(\bm{A}^n)^{(k_{1}-1,2)}_{\vec{r}+n\vec{e}_{1}}$
\begin{align}
    \text{from }(\bm{A}^n)^{(k_{1}-1,2)}_{\vec{r}+n\vec{e}_{1}}:\qquad 
    \begin{array}{|cc|}\hline
    \Igray{Z_{\vec{r}+n\vec{e}_{1}}}& \\
    (X)^{k_{1}-3-n}_{\vec{r}+(n+1)\vec{e}_{1},\vec{e}_{1}}& \\
    Y&Z\\
    (X)^{n-1}_{\vec{r}+(k_{1}-1)\vec{e}_{1},\vec{e}_{1}}& \\
    Y& \\\hline
    \end{array},
\end{align}
and one from the $(k_{1}-1,2)$-support input
$(\bm{B}^{n})^{(k_{1}-1,2)}_{\vec{r}+n\vec{e}_{1}}$
\begin{align}
    \text{from }(\bm{B}^{n})^{(k_{1}-1,2)}_{\vec{r}+n\vec{e}_{1}}:\qquad 
    \begin{array}{|cc|}\hline
    Y_{\vec{r}+n\vec{e}_{1}}& \\
    (X)^{k_{1}-3-n}_{\vec{r}+(n+1)\vec{e}_{1},\vec{e}_{1}}& \\
    Y&Z\\
    (X)^{n-1}_{\vec{r}+(k_{1}-1)\vec{e}_{1},\vec{e}_{1}}& \\
    \Igray{Z}& \\\hline
    \end{array}.
\end{align}
For this output, 
Eq.~\eqref{eq:ConservCond} reduces to
\begin{align}
    -J^{2} c_{\bm{C},\vec{r}+n\vec{e}_{1}}^{(k_{1}-1,2)}
    +h^{x} c_{(\bm{A}^n),\vec{r}+n\vec{e}_{1}}^{(k_{1}-1,2)}& \notag\\
    +h^{x}c_{(\bm{B}^{n}),\vec{r}+n\vec{e}_{1}}^{(k_{1}-1,2)}
    &=0,
    \label{eq:c_Z(X)Z_k-1_Output_2}
\end{align}
where $n\in\{1,2,...,k_{1}-3\}$.
By a similar calculation, we also have
\begin{align}
    J^{2} c_{\bm{C},\vec{r}+(k_{1}-2)\vec{e}_{1}}^{(k_{1}-1,2)}
    +h^{x}c_{(\bm{B}^{k_{1}-2}),\vec{r}+(k_{1}-2)\vec{e}_{1}}^{(k_{1}-1,2)}
    &=0.
    \label{eq:c_Z(X)Z_k-1_Output_3}
\end{align}

From Eqs.~\eqref{eq:prop_InputZZ_Zone_2}, \eqref{eq:c_Z(X)Z_Y(X)Y}, \eqref{eq:c_Z(X)Z_k_Output_1}, and \eqref{eq:c_Z(X)Z_k-1_Output_1},
we have
\begin{align}
    J^{1}c_{(\bm{B}^1),\vec{r}+\vec{e}_{1}}^{(k_{1}-1,2)}
    =2J^{2} c_{Z(X)^{k_{1}-2}Z,\vec{r}}^{(k_{1},1)}.
\end{align}
From Eqs.~\eqref{eq:prop_InputZZ_Zone_2}, \eqref{eq:c_Z(X)Z_Y(X)Y}, \eqref{eq:c_Z(X)Z_k_Output_2}, and \eqref{eq:c_Z(X)Z_k-1_Output_2},
we have
\begin{align}
    J^{1}c_{(\bm{B}^{n+1}),\vec{r}+(n+1)\vec{e}_{1}}^{(k_{1}-1,2)}
    -J^{1}c_{(\bm{B}^{n}),\vec{r}+n\vec{e}_{1}}^{(k_{1}-1,2)}
    =2J^{2} c_{Z(X)^{k_{1}-2}Z,\vec{r}}^{(k_{1},1)}
\end{align}
for $n\in\{1,2,...,k_{1}-4\}$.
From Eqs.~\eqref{eq:prop_InputZZ_Zone_2}, \eqref{eq:c_Z(X)Z_Y(X)Y}, \eqref{eq:c_Z(X)Z_k_Output_3}, and \eqref{eq:c_Z(X)Z_k-1_Output_2},
we have
\begin{align}
    &-J^{1}c_{(\bm{B}^{k_{1}-2}),\vec{r}+(k_{1}-2)\vec{e}_{1}}^{(k_{1}-1,2)}
    -J^{1}c_{(\bm{B}^{k_{1}-3}),\vec{r}+(k_{1}-3)\vec{e}_{1}}^{(k_{1}-1,2)}\notag\\
    &\quad =2J^{2} c_{Z(X)^{k_{1}-2}Z,\vec{r}}^{(k_{1},1)}
\end{align}
From Eqs.~\eqref{eq:prop_InputZZ_Zone_2}, \eqref{eq:c_Z(X)Z_Y(X)Y}, and \eqref{eq:c_Z(X)Z_k-1_Output_3},
we have
\begin{align}
    &J^{1}c_{(\bm{B}^{k_{1}-2}),\vec{r}+(k_{1}-2)\vec{e}_{1}}^{(k_{1}-1,2)}
    =J^{2} c_{Z(X)^{k_{1}-2}Z,\vec{r}}^{(k_{1},1)}.
\end{align}
Combining these results, we obtain
\begin{align}
    0&=J^{1}c_{(\bm{B}^1),\vec{r}+\vec{e}_{1}}^{(k_{1}-1,2)}\notag\\
    &\quad+J^{1}\sum_{n=1}^{k_{1}-4}\bigl(c_{(\bm{B}^{n+1}),\vec{r}+(n+1)\vec{e}_{1}}^{(k_{1}-1,2)}
    -c_{(\bm{B}^{n}),\vec{r}+n\vec{e}_{1}}^{(k_{1}-1,2)}\bigr)\notag\\
    &\quad -J^{1}c_{(\bm{B}^{k_{1}-2}),\vec{r}+(k_{1}-2)\vec{e}_{1}}^{(k_{1}-1,2)}
    -J^{1}c_{(\bm{B}^{k_{1}-3}),\vec{r}+(k_{1}-3)\vec{e}_{1}}^{(k_{1}-1,2)}\notag\\
    &\quad+J^{1}c_{(\bm{B}^{k_{1}-2}),\vec{r}+(k_{1}-2)\vec{e}_{1}}^{(k_{1}-1,2)}\notag\\
    &=(2k_{1}-3)J^{2} c_{Z(X)^{k_{1}-2}Y,\vec{r}}^{(k_{1},1)}.
\end{align}
(Note that we can obtain the same equality even when $k_{1}=3,4$.)
Thus, we have the following proposition.
\begin{proposition}[inputs $Z(X)^{k_{1}-2}Z$]\label{prop:input_ZZ_Z(X)Z}
Assume $J^{1},h^{x},J^{2}\neq 0$.
For $3\le k_{1}\le L/2$, 
any solution of Eq.~\eqref{eq:ConservCond} must satisfy 
\begin{align}
    c_{Z(X)^{k_{1}-2}Z,\vec{r}}^{(k_{1},1)}
    =0,
    \label{eq:prop_InputZZ_Z(X)Z}
\end{align}
where $c_{Z(X)^{k_{1}-2}Z,\vec{r}}^{(k_{1},1)}$ represents the coefficients corresponding to the input given by the RHS of Eq.~\eqref{eq:prop_InputZZ_Zone_1}.
\end{proposition}

Propositions~\ref{prop:input_XX}--\ref{prop:input_ZZ_Z(X)Z} prove that,
in order for any candidate of $(k_{1},k_{2})$-local conserved quantities~\eqref{eq:Q_Expansion} to satisfy Eq.~\eqref{eq:ConservCond}, 
all coefficients $c_{\bm{A},\vec{r}}^{(k_{1},\ell_{2})}$
corresponding to $(k_{1},\ell_{2})$-support inputs ($3\le k_{1}\le L/2$) must be zero,
which indicates that $k_{1}$ can be taken smaller.
In other words, $(k_{1},k_{2})$-local conserved quantities are absent 
for $3\le k_{1}\le L/2$.
Therefore, we obtain the following theorem.
\begin{theorem}\label{thm:3localConserved}
Assume $J^{1},h^{x},J^{2}\neq 0$.
For $3\le k_{1}\le L/2$, 
there are no $(k_{1},k_{2})$-local conserved quantities.
\end{theorem}
\noindent
Note that $k_{2}\in\{1,2,..., L\}$ can be taken arbitrarily 
since the above proof has imposed no restriction on the locality in the direction $\vec{e}_{2}$.
Note also that the value of $h^{z}\in\mathbb{R}$ can be taken arbitrarily, including $h^{z}=0$,
as can be seen from the fact that no diagrams containing dotted rectangles (which represent the Hamiltonian $Z$ terms) have appeared in the above proof.

\subsection{\label{sec:2D_k_le_2}Small locality case (\texorpdfstring{$k_{1}\le 2$}{k1 le 2})}

In Secs.~\ref{sec:2D_XX}--\ref{sec:2D_ZZ}, we have proved that there are no $(k_{1},k_{2})$-local conserved quantities for $3\le k_{1}\le L/2$.
In this subsection, we investigate $(k_{1},k_{2})$-local conserved quantities with $k_{1}\le 2$.

We start by considering a candidate of $(2,k_{2})$-local conserved quantities.
As mentioned in the beginning of Sec.~\ref{sec:2D}, we can classify $(2,\ell_{2})$-support inputs into three types: types~\ref{enum:input_XX}, \ref{enum:input_XZ}, and \ref{enum:input_ZZ}.
As explained in Sec.~\ref{sec:2D_XZ} (resp. Sec.~\ref{sec:2D_ZZ}),
type-\ref{enum:input_XZ} (resp. type-\ref{enum:input_ZZ}) inputs can be further divided into types~\ref{enum:input_XZ_Zmulti} and \ref{enum:input_XZ_Zone}
(resp. types~\ref{enum:input_ZZ_Zmulti} and \ref{enum:input_ZZ_Zone}).
From propositions~\ref{prop:input_XX} and \ref{prop:input_XZ_Zmulti}, 
we can show that coefficients corresponding to type-\ref{enum:input_XX} and \ref{enum:input_XZ_Zmulti} inputs must be zero.
From lemma~\ref{lemma:inputXZ_Zone_X}, the remaining $(2,\ell_{2})$-support inputs of type~\ref{enum:input_XZ_Zone} 
do not include any $X$
and take either the following form
\begin{align}
    &\begin{array}{|cccc|}\hline
    I_{\vec{r}}& &Z_{\vec{p}}& \\
    \ast&Y_{\vec{q}}&A_{\vec{p}+\vec{e}_{1}}&\ast\\\hline
    \end{array}
    \quad\text{with }A_{\vec{p}+\vec{e}_{1}}=Z,I,
    \label{eq:Example_k=2_ZY_1}\\
    &\begin{array}{|cccc|}\hline
    I_{\vec{r}}& &Z_{\vec{p}}& \\
    \ast&\ast&Y_{\vec{p}+\vec{e}_{1}}&\ast\\\hline
    \end{array},
    \label{eq:Example_k=2_ZY_2}
\end{align}
or the inputs obtained by interchanging the upper and the lower edges in Eq.~\eqref{eq:Example_k=2_ZY_1} or \eqref{eq:Example_k=2_ZY_2}.
For the former inputs~\eqref{eq:Example_k=2_ZY_1}, 
$c_{\bm{A},\vec{r}}^{(2,\ell_{2})}=0$
follows from the following diagram
\begin{align}
    \begin{array}{cccc}\hline
    \Ic{I_{\vec{r}}}& &Z_{\vec{p}}&\cI{} \\\ucline{2-2}
    \Ic{\ast}&\IIcII{Y_{\vec{q}}}& A_{\vec{p}+\vec{e}_{1}} &\cI{\ast}\\\hline
    &\IIcII{}&&\\\dcline{2-2}
    \end{array}
    \ \to \
    \begin{array}{|cccc|}\hline
    I_{\vec{r}}& &Z_{\vec{p}}& \\
    \ast&X_{\vec{q}}& A_{\vec{p}+\vec{e}_{1}} &\ast\\
    &Z&&\\\hline
    \end{array}.
\end{align}
For the latter inputs~\eqref{eq:Example_k=2_ZY_2}, we consider the following diagram
\begin{align}
    \begin{array}{|cccc|}\hline
    I_{\vec{r}}& &\gray{Z_{\vec{p}}}& \\
    \ast&\ast& Y_{\vec{p}+\vec{e}_{1}} &\ast\\\hline
    \end{array}
    \ \to \
    \begin{array}{|cccc|}\hline
    I_{\vec{r}}& &Y_{\vec{p}}& \\
    \ast&\ast& Y_{\vec{p}+\vec{e}_{1}} &\ast\\\hline
    \end{array}.
    \label{eq:diag_k=2_ZY}
\end{align}
This output contains the other contribution
\begin{align}
    \begin{array}{|c|}\hline
    Y_{\vec{p}}\\
    \IgrayI{Z_{\vec{p}+\vec{e}_{1}}}\\\hline
    \end{array}
    \ \to \
    \begin{array}{|c|}\hline
    Y_{\vec{p}}\\
    Y_{\vec{p}+\vec{e}_{1}}\\\hline
    \end{array}
\end{align}
iff. the lower edge of the input in Eq.~\eqref{eq:diag_k=2_ZY} includes only a single Pauli operator $Y_{\vec{p}+\vec{e}_{1}}$.
Thus, the remaining $(2,\ell_{2})$-support inputs of type~\ref{enum:input_XZ_Zone} are given by $Z_{\vec{r}}Y_{\vec{r}+\vec{e}_{1}}$ or $Y_{\vec{r}}Z_{\vec{r}+\vec{e}_{1}}$.

Now we consider some outputs that contain contributions from the input $Z_{\vec{r}}Y_{\vec{r}+\vec{e}_{1}}$ or $Y_{\vec{r}}Z_{\vec{r}+\vec{e}_{1}}$.
The first output considered here is the following $(3,1)$-support one
\begin{align}
    \begin{array}{|c|}\hline
    Z_{\vec{r}}\\
    X_{\vec{r}+\vec{e}_{1}}\\
    Z\\\hline
    \end{array}.
\end{align}
For this output, Eq.~\eqref{eq:ConservCond} reduces to
\begin{align}
    J^{1} c_{Z_{\vec{r}}Y_{\vec{r}+\vec{e}_{1}}}^{(2,1)}
    +J^{1} c_{Y_{\vec{r}+\vec{e}_{1}}Z_{\vec{r}+2\vec{e}_{1}}}^{(2,1)}
    =0.
    \label{eq:c_k=2_ZY_1}
\end{align}
The second output considered here is the following $(2,2)$-support one
\begin{align}
    \begin{array}{|cc|}\hline
    Z_{\vec{r}}& \\
    X_{\vec{r}+\vec{e}_{1}}&Z\\\hline
    \end{array}.
\end{align}
For this output, Eq.~\eqref{eq:ConservCond} reduces to
\begin{align}
    J^{2} c_{Z_{\vec{r}}Y_{\vec{r}+\vec{e}_{1}}}^{(2,1)}
    +J^{1} c_{Y_{\vec{r}+\vec{e}_{1}}Z_{\vec{r}+\vec{e}_{1}+\vec{e}_{2}}}^{(1,2)}
    =0.
    \label{eq:c_k=2_ZY_2}
\end{align}
The third output considered here is the following $(2,2)$-support one
\begin{align}
    \begin{array}{|cc|}\hline
    X_{\vec{r}+\vec{e}_{1}}&Z\\
    Z_{\vec{r}+2\vec{e}_{1}}& \\\hline
    \end{array}.
\end{align}
For this output, Eq.~\eqref{eq:ConservCond} reduces to
\begin{align}
    J^{2} c_{Y_{\vec{r}+\vec{e}_{1}}Z_{\vec{r}+2\vec{e}_{1}}}^{(2,1)}
    +J^{1} c_{Y_{\vec{r}+\vec{e}_{1}}Z_{\vec{r}+\vec{e}_{1}+\vec{e}_{2}}}^{(1,2)}
    =0.
    \label{eq:c_k=2_ZY_3}
\end{align}
Combining Eqs.~\eqref{eq:c_k=2_ZY_1}, \eqref{eq:c_k=2_ZY_2}, and \eqref{eq:c_k=2_ZY_3},
we have 
\begin{align}
    c_{Z_{\vec{r}}Y_{\vec{r}+\vec{e}_{1}}}^{(2,1)}
    =c_{Y_{\vec{r}+\vec{e}_{1}}Z_{\vec{r}+2\vec{e}_{1}}}^{(2,1)}
    =0.
\end{align}
Therefore, the remaining $(2,\ell_{2})$-support inputs are only of type~\ref{enum:input_ZZ}.

Next we investigate $(2,\ell_{2})$-support inputs of type~\ref{enum:input_ZZ}.
For $(2,\ell_{2})$-support inputs of type~\ref{enum:input_ZZ_Zmulti}, we consider the following diagram
\begin{align}
    \begin{array}{|cccc|}\hline
    Z_{\vec{r}}&A_{\vec{q}-\vec{e}_{1}}&Z_{\vec{p}}& \\
    \ast&\gray{Z_{\vec{q}}}&\ast &\ast\\\hline
    \end{array}
    \ \to \
    \begin{array}{|cccc|}\hline
    Z_{\vec{r}}&A_{\vec{q}-\vec{e}_{1}}&Z_{\vec{p}}& \\
    \ast&Y_{\vec{q}}&\ast &\ast\\\hline
    \end{array}.
\end{align}
By the same reason explained below Eq.~\eqref{eq:diag_InputZZ_Zmulti} of Sec.~\ref{sec:2D_ZZ}, there are no other contributions, 
and Eq.~\eqref{eq:ConservCond} reduces to
\begin{align}
    h^{x} c_{\bm{A},\vec{r}}^{(2,\ell_{2})}=0\quad \text{for all }\bm{A}^{(2,\ell_{2})}_{\vec{r}}\text{ of type \ref{enum:input_ZZ_Zmulti}}.
\end{align}

For $(2,\ell_{2})$-support inputs of type~\ref{enum:input_ZZ_Zone}, we consider the following diagram
\begin{align}
    \begin{array}{|cccc|}\hline
    I_{\vec{r}}& &\gray{Z_{\vec{p}}}& \\
     & Z_{\vec{q}} & & \\\hline
    \end{array}
    \ \to \
    \begin{array}{|cccc|}\hline
    I_{\vec{r}}& &Y_{\vec{p}}& \\
     & Z_{\vec{q}} & & \\\hline
    \end{array}.
    \label{eq:diag_k=2_ZZ,X}
\end{align}
This output contains the other contribution
\begin{align}
    \begin{array}{c}\ucline{1-1}
    \IIcII{X_{\vec{p}}}\\\hline
    \IIcII{\quad}\\\dcline{1-1}
    \end{array}
    \ \to \
    \begin{array}{|c|}\hline
    Y_{\vec{p}}\\
    Z_{\vec{p}+\vec{e}_{1}}\\\hline
    \end{array}
    \label{eq:diag_k=2_X,ZZ1}
\end{align}
iff. $\vec{q}=\vec{p}+\vec{e}_{1}$ holds.
Thus, the remaining $(2,\ell_{2})$-support inputs are restricted to $Z_{\vec{r}}Z_{\vec{r}+\vec{e}_{1}}$, which is originally included in the Hamiltonian.

Now we investigate $(1,\ell_{2})$-support inputs [for a candidate of $(2,k_{2})$-local conserved quantities]. 
Such inputs can be classified into two types:
input that includes some non-$Z$ Pauli operators ($X$ or $Y$)
and input that consists of only $Z$ or $I$.
For the former inputs, we consider the following diagram
\begin{align}
    \begin{array}{cccc}\ucline{3-3}\hline
    \Ic{A_{\vec{r}}}&\ast&\IIcII{X_{\vec{p}}}&\cI{\ast}\\\hline
     & &\IIcII{}& \\\dcline{3-3}
    \end{array}
    \ \to \
    \begin{array}{|cccc|}\hline
    A_{\vec{r}}&\ast&Y_{\vec{p}}&\ast\\
     & &Z& \\\hline
    \end{array}.
\end{align}
This output contains no other contributions unless the above input is exactly the same as that of Eq.~\eqref{eq:diag_k=2_X,ZZ1}.
Thus, from Eq.~\eqref{eq:ConservCond}, we have
\begin{align}
    c_{\bm{A},\vec{r}}^{(1,\ell_{2})}=0\quad &\text{if }A_{\vec{p}}=X,Y\text{ for some }\vec{p}\notag\\
    &\text{unless }\bm{A}^{(1,\ell_{2})}_{\vec{r}}\text{ is not a single }X_{\vec{p}}.
    \label{eq:c_k=1_InputX}
\end{align}
Therefore the remaining $(1,\ell_{2})$-support inputs are
inputs that consist of a single $X$
and inputs that consist of only $Z$ or $I$.
The latter inputs can be written as 
\begin{align}
    \begin{array}{|cccc|}\hline
    A_{\vec{r}}&\ast&Z_{\vec{p}}&\ast\\\hline
    \end{array}.
\end{align}
For such inputs,
we consider the following diagram
\begin{align}
    \begin{array}{|cccc|}\hline
    A_{\vec{r}}&\ast&\gray{Z_{\vec{p}}}&\ast\\\hline
    \end{array}
    \ \to \
    \begin{array}{|cccc|}\hline
    A_{\vec{r}}&\ast&Y_{\vec{p}}&\ast\\\hline
    \end{array}.
    \label{eq:diag_k=1_onlyZ,X}
\end{align}
Since the remaining $(2,\ell_{2})$-support and $(1,\ell_{2})$-support inputs satisfy $A_{\vec{p}}=X,Z,I$, we need to apply a certain Hamiltonian term to the site $\vec{p}$ in order to obtain $Y_{\vec{p}}\, (\neq A_{\vec{p}})$ in the above output.
Hence, the other contributions to the above output are restricted to the following
\begin{align}
    \begin{array}{p{5cm}c}\ucline{1-2}
    \IIc{\qquad}&\IcII{X_{\vec{p}}}\\\dcline{1-2}
    \end{array}
    \ &\to \
    \begin{array}{|cc|}\hline
    Z_{\vec{p}-\vec{e}_{2}}&Y_{\vec{p}}\\\hline
    \end{array},
    \label{eq:diag_k=1_X,ZZ_1}\\
    \begin{array}{cc}\ucline{1-2}
    \IIc{X_{\vec{p}}}&\IcII{\quad}\\\dcline{1-2}
    \end{array}
    \ &\to \
    \begin{array}{|cc|}\hline
    Y_{\vec{p}}&Z_{\vec{p}+\vec{e}_{2}}\\\hline
    \end{array},
    \label{eq:diag_k=1_X,ZZ_2}\\
    \begin{array}{c}\ucline{1-1}
    \IIcII{X_{\vec{p}}}\\\ucline{1-1}
    \end{array}
    \ &\to \
    \begin{array}{|c|}\hline
     Y_{\vec{p}}\\\hline
    \end{array}.
    \label{eq:diag_k=1_X,Z_1}
\end{align}
[If the output of Eq.~\eqref{eq:diag_k=1_onlyZ,X} does not coincide with either the output of Eq.~\eqref{eq:diag_k=1_X,ZZ_1}, \eqref{eq:diag_k=1_X,ZZ_2}, or \eqref{eq:diag_k=1_X,Z_1}, then there are no other contributions. 
In this case, $h^{x} c_{\bm{A},\vec{r}}^{(1,\ell_{2})}=0$ for the input $\bm{A}^{(1,\ell_{2})}_{\vec{r}}$ in Eq.~\eqref{eq:diag_k=1_onlyZ,X} follows from Eq.~\eqref{eq:ConservCond} for this output.]
This means that all the remaining inputs are given by 
$\bm{A}^{(\ell_{1},\ell_{2})}_{\vec{r}}=Z_{\vec{r}}Z_{\vec{r}+\vec{e}_{1}}$, $X_{\vec{r}}$, $Z_{\vec{r}}Z_{\vec{r}+\vec{e}_{2}}$, and $Z_{\vec{r}}$.
In addition, the above diagrams~\eqref{eq:diag_k=1_onlyZ,X}--\eqref{eq:diag_k=1_X,ZZ_2} also imply that 
\begin{align}
    J^{2}c_{X_{\vec{r}}}^{(1,1)}
    =h^{x}c_{Z_{\vec{r}}Z_{\vec{r}+\vec{e}_{2}}}^{(1,2)}
    =J^{2}c_{X_{\vec{r}+\vec{e}_{2}}}^{(1,1)},
    \label{eq:c_k=1_X_ZZ_e2}
\end{align}
and the diagrams~\eqref{eq:diag_k=1_onlyZ,X} and \eqref{eq:diag_k=1_X,Z_1} imply that 
\begin{align}
    h^{z}c_{X_{\vec{r}}}^{(1,1)}
    =h^{x}c_{Z_{\vec{r}}}^{(1,1)}.
    \label{eq:c_k=1_X_Z}
\end{align}
Furthermore, from Eqs.~\eqref{eq:diag_k=2_ZZ,X} and \eqref{eq:diag_k=2_X,ZZ1}, we have
\begin{align}
    J^{1}c_{X_{\vec{r}}}^{(1,1)}
    =h^{x}c_{Z_{\vec{r}}Z_{\vec{r}+\vec{e}_{1}}}^{(2,1)}.
    \label{eq:c_k=1_X_ZZ_e1_1}
\end{align}
In addition, by considering the following $(2,1)$-support output,
\begin{align}
    \begin{array}{|c|}\hline
    Z_{\vec{r}}\\
    Y_{\vec{r}+\vec{e}_{1}}\\\hline
    \end{array},
\end{align}
we have
\begin{align}
    J^{1}c_{X_{\vec{r}+\vec{e}_{1}}}^{(1,1)}
    =h^{x}c_{Z_{\vec{r}}Z_{\vec{r}+\vec{e}_{1}}}^{(2,1)}.
    \label{eq:c_k=1_X_ZZ_e1_2}
\end{align}
From these results, we obtain the following
\begin{align}
    &c_{\bm{A},\vec{r}}^{(\ell_{1},\ell_{2})}=0\quad
    \text{unless }\bm{A}^{(\ell_{1},\ell_{2})}_{\vec{r}}=X_{\vec{r}}, Z_{\vec{r}}, Z_{\vec{r}}Z_{\vec{r}+\vec{e}_{1}}, Z_{\vec{r}}Z_{\vec{r}+\vec{e}_{2}},
    \label{eq:c_k=2_=0}\\
    &c_{X_{\vec{r}}}^{(1,1)},\, c_{Z_{\vec{r}}}^{(1,1)},\, c_{Z_{\vec{r}}Z_{\vec{r}+\vec{e}_{1}}}^{(2,1)},\, c_{Z_{\vec{r}}Z_{\vec{r}+\vec{e}_{2}}}^{(1,2)}=\text{const. indep. of }\vec{r}
    \label{eq:c_k=2_SiteIndependence}\\
    &c_{X_{\vec{r}}}^{(1,1)}: c_{Z_{\vec{r}}}^{(1,1)}: c_{Z_{\vec{r}}Z_{\vec{r}+\vec{e}_{1}}}^{(2,1)}: c_{Z_{\vec{r}}Z_{\vec{r}+\vec{e}_{2}}}^{(1,2)}=h^{x}: h^{z}: J^{1}: J^{2}
    \label{eq:c_k=2_propto_H}
\end{align}
These show that there are no $(2,k_{2})$-local conserved quantities other than (a linear combination of the identity and) the Hamiltonian.

For completeness, we comment on a candidate of $(1,k_{2})$-local conserved quantities.
Because the above analysis of a candidate of $(2,k_{2})$-local conserved quantities
applies also to a candidate of $(1,k_{2})$-local ones, 
all the above equations are valid with the coefficients of $(2,\ell_{2})$-support inputs set to zero.
Therefore, from Eqs.~\eqref{eq:c_k=2_=0} and \eqref{eq:c_k=2_propto_H}, 
we can see that all coefficients $c_{\bm{A},\vec{r}}^{(1,\ell_{2})}$ become zero.
This means that there are no $(1,k_{2})$-local conserved quantities.

Combining these, we obtain the following theorem.
\begin{theorem}\label{thm:2localConserved}
Assume $J^{1},h^{x},J^{2}\neq 0$.
For $k_{1}\le 2$, 
any $(k_{1},k_{2})$-local conserved quantity is restricted to
a linear combination of the Hamiltonian and the identity 
[i.e., a $(2,2)$-local one].
\end{theorem}
\noindent
Note that, as in Theorem~\ref{thm:3localConserved}, $k_{2}\in\{1,2,...,L\}$ and the value of $h^{z}\in\mathbb{R}$ can be taken arbitrarily, including $h^{z}=0$.

\section{\label{sec:3D}Proof in higher dimension}

In this section, we explain how the proof of the previous section, Sec.~\ref{sec:2D}, can be extended to the model~\eqref{eq:Hamiltonian} with $d>2$.

First we define ``faces'' instead of the edges $E^{(\ell_{1},\ell_{2})}_{\bm{A},\vec{r}}$ and $F^{(\ell_{1},\ell_{2})}_{\bm{A},\vec{r}}$ in Sec.~\ref{sec:2D}. 
Without loss of generality, we take  $\mu^*=1$, that is, we assume that $k_{1}\le L/2$. 
Take an arbitrary Pauli product $\bm{A}^{(\ell_{1},...,\ell_{d})}_{\vec{r}}$.
The following two faces of the rectangle $R^{(\ell_{1},...,\ell_{d})}_{\vec{r}}$ play a crucial role in the proof,
\begin{align}
    E^{(\ell_{1},...,\ell_{d})}_{\bm{A},\vec{r}}&=\bigl\{(r_{1},p_{2},...,p_{d})\in\Lambda \bigm| \notag\\
    &\qquad r_{\mu}\le p_{\mu}\le r_{\mu}+\ell_{\mu}-1\text{ for }\mu=2,...,d\bigr\}\\
    F^{(\ell_{1},...,\ell_{d})}_{\bm{A},\vec{r}}&=\{(r_{1}+\ell_{1}-1,p_{2},...,p_{d})\in\Lambda \bigm|\notag\\
    &\qquad r_{\mu}\le p_{\mu}\le r_{\mu}+\ell_{\mu}-1\text{ for }\mu=2,...,d\bigr\},
\end{align}
where $\vec{r}=(r_{1},r_{2},...,r_{d})$.
For any Pauli product $\bm{A}^{(\ell_{1},\ell_{2})}_{\vec{r}}$, there is at least one Pauli operator $A_{\vec{p}}\neq I_{\vec{p}}$ on each face, as explained below Eq.~\eqref{eq:PauliProd_A}.

Now we explain that
the proof in Sec.~\ref{sec:2D} holds in almost the same way, 
just by replacing 
the edges $E^{(\ell_{1},\ell_{2})}_{\bm{A},\vec{r}}$ and $F^{(\ell_{1},\ell_{2})}_{\bm{A},\vec{r}}$ 
with the faces $E^{(\ell_{1},...,\ell_{d})}_{\bm{A},\vec{r}}$ and $F^{(\ell_{1},...,\ell_{d})}_{\bm{A},\vec{r}}$.
As explained in Secs.~\ref{sec:2D}, 
$(k_{1},\ell_{2},...,\ell_{d})$-support inputs $\bm{A}^{(k_{1},\ell_{2},...,\ell_{d})}_{\vec{r}}$
can be classified into the following types:
\begin{enumerate}[label={\roman*.},ref={\roman*},align=right,leftmargin=!]
    \item \label{enum:input_3D_XX}An input 
    that includes $X$ or $Y$ on both faces
    \item \label{enum:input_3D_XZ}An input 
    that includes $X$ or $Y$ on one face but not on the other face
    \begin{enumerate}[label={\ref{enum:input_3D_XZ}-\alph*.},ref={\ref{enum:input_XZ}-\alph*},align=right,leftmargin=!]
        \item \label{enum:input_3D_XZ_Zmulti}A type-\ref{enum:input_3D_XZ} input that includes more than one $Z$ on its $Z$ edge
        \item \label{enum:input_3D_XZ_Zone}A type-\ref{enum:input_3D_XZ} input that includes exactly one $Z$ on its $Z$ edge
    \end{enumerate}
    \item \label{enum:input_3D_ZZ}An input 
    that does not include $X$ nor $Y$ on both faces
    \begin{enumerate}[label={\ref{enum:input_3D_ZZ}-\alph*.},ref={\ref{enum:input_3D_ZZ}-\alph*},align=right,leftmargin=!]
        \item\label{enum:input_3D_ZZ_Zmulti}A type-\ref{enum:input_3D_ZZ} input that includes more than one $Z$ on either face
        \item\label{enum:input_3D_ZZ_Zone}A type-\ref{enum:input_3D_ZZ} input that includes exactly one $Z$ on both faces
    \end{enumerate}
\end{enumerate}
Under these classifications, Propositions~\ref{prop:input_XX}--\ref{prop:input_XZ_Zone} hold without any major changes.

For the proof of Proposition~\ref{prop:input_XZ_Z(X)Y} in $d=2$, 
we have focused on a series of outputs~\eqref{eq:Output_Z(X)Y_1}, \eqref{eq:Output_Z(X)Y_2}, \eqref{eq:Output_Z(X)Y_3}, \eqref{eq:Output_Z(X)Y_4}, \eqref{eq:Output_Z(X)Y_5}, and \eqref{eq:Output_Z(X)Y_6}. 
This proof is applicable even in $d>2$
by considering the corresponding outputs on a plane spanned by $\vec{e}_{1}$ and $\vec{e}_{2}$.
For instance, we consider the $(k_{1},2,1,...,1)$-support output
\begin{align}
    Z_{\vec{r}}\Bigl(\prod_{n=1}^{k_{1}-1}X_{\vec{r}+n\vec{e}_{1}}\Bigr)Z_{\vec{r}+(k_{1}-1)\vec{e}_{1}+\vec{e}_{2}},
\end{align}
corresponding to Eq.~\eqref{eq:Output_Z(X)Y_1}.
All the inputs and outputs appearing in such a proof are supported on that plane, and hence the proof does not contain the parameters $J^{3},..., J^{d}$.
These mean that Proposition~\ref{prop:input_XZ_Z(X)Y} holds even in $d>2$ under the assumption  $J^{1},h^{x},J^{2}\neq 0$,
and it is not necessary to assume $J^{\mu}\neq 0 $ for $\mu=3,...,d$.

The proof of Propositions~\ref{prop:input_ZZ_Zmulti} and \ref{prop:input_ZZ_Zone} needs no major changes.
The proof of Proposition~\ref{prop:input_ZZ_Z(X)Z} can be extended to $d>2$ in the same manner as that of Proposition~\ref{prop:input_XZ_Z(X)Y}.
As a result, we obtain the following theorem.

\begin{theorem}\label{thm:3D_3localConserved}
Assume $d\ge 2$, $J^{1},h^{x},J^{2}\neq 0$ while $J^{3},...,J^{d}$, $h^{z}$ are arbitrary.
For $3\le k_{1}\le L/2$, 
there are no $(k_{1},k_{2},...,k_{d})$-local conserved quantities.
\end{theorem}
\noindent
It seems surprising that 
the assumptions $J^{\mu}\neq 0$ for all $\mu=1,..,d$ are not necessary 
but only those for two of them 
(in the above theorem, we take $J^{1}$ and $J^{2}$) 
are sufficient for 
the absence of $(k_{1},k_{2},...,k_{d})$-local conserved quantities with $3\le k_{1}\le L/2$.
In other words, 
even if $J^{3}=...=J^{d}=0$, 
the Hamiltonian~$H$ 
(which can be written as an array of $L^{d-2}$ number of two dimensional Hamiltonians~$H_{\mathrm{2D}}$ in such a case) 
possesses no $(k_{1},k_{2},...,k_{d})$-local conserved quantities with $3\le k_{1}\le L/2$.
Note that, although in such a system $H_{\mathrm{2D}}$ and its polynomials become trivial conserved quantities, this fact does not contradict the above theorem 
because $H_{\mathrm{2D}}$ is a $(2,2,1,...,1)$-local conserved quantity 
and $(H_{\mathrm{2D}})^2,(H_{\mathrm{2D}})^3,(H_{\mathrm{2D}})^4,...$ are $(k_{1},k_{2},1,...,1)$-local conserved quantities with $k_{1},k_{2}>L/2$, 
both of which are out of the scope of the theorem.

For the proof of Theorem~\ref{thm:2localConserved}, the analysis of $(2,\ell_{2})$-support inputs and $(1,\ell_{2})$-support inputs that include some non-$Z$ Pauli operators, Eqs.~\eqref{eq:Example_k=2_ZY_1}--\eqref{eq:c_k=1_InputX}, needs no major change.
On the other hand, 
in the analysis of $(1,\ell_{2})$-support inputs that consist of only $Z$ or $I$, 
Eqs.~\eqref{eq:diag_k=1_X,ZZ_1} and \eqref{eq:diag_k=1_X,ZZ_2} need to be modified as
\begin{align}
    &[c_{X_{\vec{p}}}^{(1,...,1)}X_{\vec{p}},
    J^{\mu}Z_{\vec{p}-\vec{e}_{\mu}}Z_{\vec{p}}]/2i\notag\\
    &\quad =-c_{X_{\vec{p}}}^{(1,...,1)}J^{\mu}
    Z_{\vec{p}-\vec{e}_{\mu}}Y_{\vec{p}}\quad (\mu=2,3,...,d),\\
    &[c_{X_{\vec{p}}}^{(1,...,1)}X_{\vec{p}},
    J^{\mu}Z_{\vec{p}}Z_{\vec{p}+\vec{e}_{\mu}}]/2i\notag\\
    &\quad =-c_{X_{\vec{p}}}^{(1,...,1)}J^{\mu}
    Y_{\vec{p}}Z_{\vec{p}+\vec{e}_{\mu}}\quad (\mu=2,3,...,d),
\end{align}
respectively.
Combining these with the contribution~\eqref{eq:diag_k=1_onlyZ,X},
we have
\begin{align}
    J^{\mu}c_{X_{\vec{r}}}^{(1,...,1)}
    =h^{x}c_{Z_{\vec{r}}Z_{\vec{r}+\vec{e}_{\mu}}}^{(1,...1,2,1,...,1)}
    =J^{\mu}c_{X_{\vec{r}+\vec{e}_{\mu}}}^{(1,...,1)}
    \ (\mu=2,...,d).
\end{align}
Note that Eqs.~\eqref{eq:c_k=1_X_Z}, \eqref{eq:c_k=1_X_ZZ_e1_1}, and \eqref{eq:c_k=1_X_ZZ_e1_2} hold without any major change.
These equations result in 
\begin{align}
    &c_{\bm{A},\vec{r}}^{(\ell_{1},...,\ell_{d})}=0\quad
    \text{unless }\bm{A}^{(\ell_{1},...,\ell_{d})}_{\vec{r}}=
    Z_{\vec{r}}Z_{\vec{r}+\vec{e}_{\mu}} (\mu=1,...,d), \notag\\
    &\hspace{150pt} X_{\vec{r}}, Z_{\vec{r}},
    \\
    &c_{X_{\vec{r}}}^{(1,...,1)},\, c_{Z_{\vec{r}}}^{(1,...,1)},\, c_{Z_{\vec{r}}Z_{\vec{r}+\vec{e}_{\mu}}}^{(1,...,1,2,1,...,1)}
    =\text{const. indep. of }\vec{r}
    \\
    &c_{X_{\vec{r}}}^{(1,...,1)}: c_{Z_{\vec{r}}}^{(1,...,1)}: c_{Z_{\vec{r}}Z_{\vec{r}+\vec{e}_{\mu}}}^{(1,...,1,2,1,...,1)}
    =h^{x}: h^{z}: J^{\mu}
    \label{eq:c_3D_k=2_propto_H}
\end{align}
This means that there are no $(2,k_{2},...,k_{d})$-local conserved quantities other than (a linear combination of the identity and) the Hamiltonian.
As explained at the end of Sec.~\ref{sec:2D_k_le_2},
the above analysis also shows that there are no $(1,k_{2},...,k_{d})$-local conserved quantities.
Combining these, we obtain the following theorem.
\begin{theorem}\label{thm:3D_2localConserved}
Assume $d\ge 2$, $J^{1},J^{2},...,J^{d},h^{x}\neq 0$ while $h^{z}$ is arbitrary.
For $k_{1}\le 2$, 
any $(k_{1},...,k_{d})$-local conserved quantity is restricted to
a linear combination of the Hamiltonian and the identity 
[i.e., a $(2,...,2)$-local one].
\end{theorem}

From Theorems~\ref{thm:3D_3localConserved} and \ref{thm:3D_2localConserved}, we can obtain the main result explained in Sec.~\ref{sec:Setup}.

\section{\label{sec:Ladder}Ladder case}

In this section, we briefly explain that the proof given in Sec.~\ref{sec:2D} applies also to the quantum Ising ladder.

We consider spin-$1/2$ systems on the ladder 
$\Lambda=\{(r_{1},r_{2})\in\mathbb{Z}^2|1\le r_{1}\le L, 1\le r_{2}\le 2\}$.
The Hamiltonian is given by
\begin{align}
    H&=\sum_{\vec{r}\in\Lambda}\Bigl(J^{1}Z_{\vec{r}}Z_{\vec{r}+\vec{e}_{1}}+h^{x}X_{\vec{r}}+h^{z}Z_{\vec{r}}\Bigr)\notag\\
    &\qquad 
    +\sum_{r_{1}=1}^{L}J^{2}Z_{(r_{1},1)}Z_{(r_{1},2)}.
    \label{eq:Hamiltonian_Ladder}
\end{align}
Here, we imposed the periodic boundary condition in the direction $\vec{e}_{1}$ and the open boundary condition in the direction $\vec{e}_{2}$.
Note that 
the case where the periodic boundary conditions are imposed in both directions $\vec{e}_{1}$ and $\vec{e}_{2}$ can be reduced to the above case,
just by doubling the value of $J^{2}$.

In such a model, we say that 
an operator commuting with $H$ is
a $k_{1}$-local conserved quantity
if it is either a $(k_{1},2)$-local or $(k_{1},1)$-local conserved quantity in the sense of $d=2$ case.

Now we explain how the proof in Sec.~\ref{sec:2D} can be extended to the above model.
Since the analysis in Sec.~\ref{sec:2D} deals with all inputs in the $d=2$ case, it also includes all inputs in the ladder case.
Therefore, to complete the proof in the ladder case, it is sufficient to check that if we start the discussion of Sec.~\ref{sec:2D} from any input defined on the ladder, then we encounter only the inputs and outputs that are also defined on the ladder.
We will check this below.

In the analysis in Secs.~\ref{sec:2D_XX}--\ref{sec:2D_ZZ} except for Secs.~\ref{sec:2D_XZ_Z(X)Y} and \ref{sec:2D_ZZ_Z(X)Z}, $J^{2}$ does not appear.
Since taking the commutator with the Hamiltonian terms other than $J^{2}$ does not change the length of the support in the direction $\vec{e}_{2}$, the outputs are contained in the ladder if the inputs are so.
This means that the discussions in these sections hold even in the ladder case.
Furthermore, inputs and outputs considered in Secs.~\ref{sec:2D_XZ_Z(X)Y} and \ref{sec:2D_ZZ_Z(X)Z} are of $(\ell_{1},\ell_{2})$-support ones with $\ell_{2}\le 2$.
Therefore, the discussions in these sections also hold in the ladder case.
For the same reasons, the discussions in Sec.~\ref{sec:2D_k_le_2} also hold in the ladder case, and we obtain the following theorem.

\begin{theorem}[Ladder case]
Suppose that the coupling constants in model~\eqref{eq:Hamiltonian_Ladder} other than $h^z$ are nonzero. 
Let $k$ be a positive integer satisfying $k\le L/2$. 
When $3\le k\le L/2$, the model has no $k$-local conserved quantities. 
Furthermore, when $k\le 2$, any $k$-local conserved quantity is restricted to a linear combination of the Hamiltonian~$H$ and the identity~$I$ [i.e., a $2$-local one].
\end{theorem}

For the relation to the previous result~\cite{VanVoorden2020}, which examined 
the effective Hamiltonian obtained by taking the weak-transverse-field limit of the transverse-field Ising ladder,
see the last paragraph of Sec.~\ref{sec:Integrability}.

\section{\label{sec:ExtensionOtherLattice}Discussion on extension to other lattices}

In this section, we discuss how our results will be extended to other types of two-dimensional lattices: 
the honeycomb lattice and the triangular lattice.

\subsection{\label{sec:Honeycomb}Honeycomb lattice}

In this subsection, we provide a simple strategy to apply our proof given in Sec.~\ref{sec:2D} to the honeycomb lattice.

\begin{figure}
    \centering
    \includegraphics[width=0.9\linewidth]{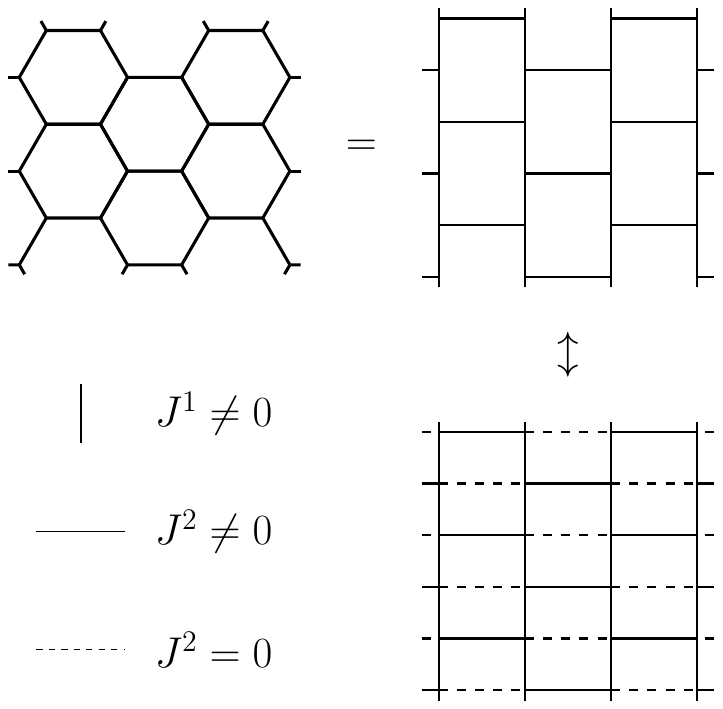}
    \caption{The honeycomb lattice (upper left) is graphically isomorphic to the brick-wall lattice (upper right). The Ising models on those lattices are equivalent to the same model on the square lattice whose Ising interactions on the edges represented by the dashed lines are absent (lower right).}
    \label{fig:Honeycomb_BrickWall}
\end{figure}

For simplicity, this subsection considers the quantum Ising model on the honeycomb lattice
whose Ising interactions on all pairs of nearest-neighbor sites take the same value.
First, we remark that 
the honeycomb lattice is graphically isomorphic to the brick-wall lattice~\cite{Chen2008,Hou2015}, as illustrated in Fig.~\ref{fig:Honeycomb_BrickWall}.
Moreover, the quantum Ising model on the brick-wall lattice can be regarded as a special case of the quantum Ising model on the square lattice where the Ising interactions $J^2$ on some nearest-neighbor sites are taken zero, as in Fig.~\ref{fig:Honeycomb_BrickWall}.
As mentioned in Sec.~\ref{sec:Ladder}, the analysis in Secs.~\ref{sec:2D_XX}--\ref{sec:2D_ZZ} except for Secs.~\ref{sec:2D_XZ_Z(X)Y} and \ref{sec:2D_ZZ_Z(X)Z} does not assume $J^{2}\neq 0$, and hence the results in these sections hold even in such models.
Furthermore, the analysis in Secs.~\ref{sec:2D_XZ_Z(X)Y} and \ref{sec:2D_ZZ_Z(X)Z} will also be valid in such models
because every site has a nearest-neighbor site in either the $\vec{e}_2$ or $-\vec{e}_2$ direction, and it is sufficient to assume $J^{2}\neq 0$ in either direction for our proof.
These indicate that in the quantum Ising model on the honeycomb lattice, Theorem~\ref{thm:3localConserved} will be proved in the same manner as in Sec.~\ref{sec:2D}.
In addition, we expect that we can extend Theorem~\ref{thm:2localConserved} in almost the same way.
These mean that our main result described in Sec.~\ref{sec:Setup} will hold even in the honeycomb lattice.

\subsection{Triangular lattice}

In this subsection, we explain difficulties in naively extending our proof given in Sec.~\ref{sec:2D} to the triangular lattice. 
We also provide a strategy to overcome some of these difficulties.

\begin{figure}
    \centering
    \includegraphics[width=\linewidth]{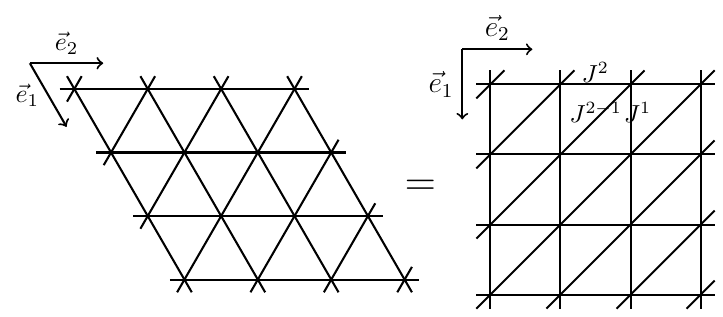}
    \caption{The triangular lattice is graphically isomorphic to the lattice obtained from the square lattice by connecting each site with its next-nearest-neighbor site separated by $\vec{e}_{2}-\vec{e}_{1}$.}
    \label{fig:Triangular_Square}
\end{figure}

For simplicity, this subsection considers the quantum Ising model on the triangular lattice
whose Ising interactions on all pairs of nearest-neighbor sites take the same value.
The quantum Ising model on the triangular lattice can be regarded as the same model on the square lattice that additionally contains the Ising interaction $J^{2-1}$ between each site and its next-nearest-neighbor site separated by $\vec{e}_{2-1}=\vec{e}_{2}-\vec{e}_{1}$, as shown in Fig.~\ref{fig:Triangular_Square}.

This additional interaction invalidates the naive extension of our proof in Sec.~\ref{sec:2D} to the triangular lattice because, for instance, the $(k_{1}+1,\ell_{2})$-support output in Eq.~\eqref{eq:diag_InputXX} of Sec.~\ref{sec:2D_XX} can contain another contribution, 
\begin{align}
    \begin{array}{ccccc}\hline
    \Ic{X_{\vec{r}}}&\ast&\ast&\ast&\cI{\ast}\\
    \Ic{\ast}&\ast&\ast&\ast&\cI{\ast}\\\ucline{4-4}\cline{4-4}
    \Ic{\ast}&\ast&Y_{\vec{p}}&\IIcII{X_{\vec{p}+\vec{e}_2}}&\cI{\ast}\\\ucline{3-3}\hline\dcline{4-4}
    &&\IIcII{}&&\\\cline{3-3}\dcline{3-3}
    \end{array}
     \ &\to \ 
    \begin{array}{|ccccc|}\hline
    X_{\vec{r}}&\ast&\ast&\ast&\ast\\
    \ast&\ast&\ast&\ast&\ast\\
    \ast&\ast&Y_{\vec{p}}&Y_{\vec{p}+\vec{e}_{2}}&\ast\\
    &&Z_{\vec{p}+\vec{e}_{1}}&&\\\hline
    \end{array},
\end{align}
where the two double-border rectangles on the LHS represent the Ising interaction $J^{2-1}$ between site $\vec{p}+\vec{e}_{1}$ and its next-nearest-neighbor site $\vec{p}+\vec{e}_{2}$.
Therefore, 
even the proof of Proposition~\ref{prop:input_XX}, the starting point of our proof, breaks down, 
and we need to change some parts of our proof.

Next, we explain that the difficulty in the proof of at least Proposition~\ref{prop:input_XX} is avoidable by a simple modification.
As explained in Sec.~\ref{sec:2D}, to prove Proposition~\ref{prop:input_XX} for the square lattice, we have applied the $ZZ$ term to an arbitrary non-$Z$ site on the lower edge $F^{(k_{1},\ell_{2})}_{\bm{A},\vec{r}}$, as in Eq.~\eqref{eq:diag_InputXX}.
In the case of the triangular lattice, we instead apply the $ZZ$ term to the rightmost non-$Z$ site $\vec{q}$ on the lower edge.
Then, the resulting $(k_{1}+1,\ell_{2})$-support output of Eq.~\eqref{eq:diag_InputXX} does not contain 
the following contribution
\begin{align}
    \begin{array}{ccccc}\hline
    \Ic{X_{\vec{r}}}&\ast&\ast&\ast&\cI{\ast}\\
    \Ic{\ast}&\ast&\ast&\ast&\cI{\ast}\\\ucline{4-4}\cline{4-4}
    \Ic{\ast}&\ast&Y_{\vec{q}}&\IIcII{A_{\vec{q}+\vec{e}_2}}&\cI{\ast}\\\ucline{3-3}\hline\dcline{4-4}
    &&\IIcII{}&&\\\cline{3-3}\dcline{3-3}
    \end{array}
     \ &\not\to \ 
    \begin{array}{|ccccc|}\hline
    X_{\vec{r}}&\ast&\ast&\ast&\ast\\
    \ast&\ast&\ast&\ast&\ast\\
    \ast&\ast&Y_{\vec{q}}&B_{\vec{q}+\vec{e}_{2}}&\ast\\
    &&Z_{\vec{q}+\vec{e}_{1}}&&\\\hline
    \end{array}
\end{align}
because $\vec{q}$ is the rightmost non-$Z$ site of the input of Eq.~\eqref{eq:diag_InputXX}, and thus $B_{\vec{q}+\vec{e}_{2}}=I,Z$.
By this modification, we can obtain Proposition~\ref{prop:input_XX} even in the triangular lattice.

We also expect that similar modifications will be possible for some of the other propositions.
However, for instance, in the proof of Lemma~\ref{lemma:inputXZ_ZoneY_Y}, the analysis of the $(k_{1}+1,\ell_{2})$-support output given by the RHS of Eq.~\eqref{eq:diag_InputXZ_Zone_Y,ZZ} will become complicated because it can contain another contribution
\begin{align}
    \begin{array}{ccccc}\ucline{4-4}
    I_{\vec{r}}&&&\IIcII{}& \\\ucline{3-3}\hline\dcline{4-4}
    \Ic{\ast}&\ast&\IIcII{A} &Y_{\vec{p}+\vec{e}_{1}} &\cI{\ast}\\\dcline{3-3}
    \Ic{\ast}&\ast&\ast&\ast&\cI{\ast}\\
    \Ic{\ast}&X&\ast&\ast&\cI{\ast}\\
    \Ic{}&Z&&&\cI{}\\\hline
    \end{array}
    \ &\to \
    \begin{array}{|ccccc|}\hline
    I_{\vec{r}}&&&Z_{\vec{p}}& \\
    \ast&\ast&B& Y_{\vec{p}+\vec{e}_{1}} &\ast\\
    \ast&\ast&\ast&\ast&\ast\\
    \ast&X&\ast&\ast&\ast\\
    &Z&&&\\\hline
    \end{array}.
\end{align}
As a result, it is no longer clear whether Lemma~\ref{lemma:inputXZ_ZoneY_Y} holds in the present form.
A detailed analysis of extending Lemma~\ref{lemma:inputXZ_ZoneY_Y} and subsequent propositions is left for future work.

\section{\label{sec:Discussion}Summary}

We have proved that conserved quantities satisfying a certain locality condition are restricted to trivial ones in the higher-dimensional quantum Ising models.
We have analyzed both zero and nonzero longitudinal field cases
on a hypercubic lattice of any dimension greater than one,
while the Ising interactions and the transverse field are taken nonzero.
Our results state that any conserved quantity that is a linear combination of operators
whose support sizes are sufficiently small 
is restricted to the Hamiltonian.
Furthermore, they impose
a strong restriction on the locality of 
conserved quantities;
any conserved quantity other than the Hamiltonian
must contain, in every spatial direction, 
at least one operator whose side length of the support in that direction 
is greater than half the linear length of the system.
These results reveal the structures of conserved quantities,
which are deeply related to the quantum nonintegrability and are consistent with the existing numerical results on level spacing statistics,
in the entire region of the parameter space of the model.

In addition, we have shown that the above result is valid even in a ladder system, where the proof structure is almost the same as that in two dimension.
We have also discussed how our results will be extended to other two-dimensional lattices, the honeycomb one and the triangular one, while a detailed analysis of such cases remains future work.

\bigskip

\textit{Note added.}
When preparing the present paper, we were informed that Shiraishi and Tasaki had independently obtained 
a proof of absence of nontrivial local conserved quantities in the spin-$1/2$ XY and XYZ models in
two or higher dimensions~\cite{Shiraishi2024a}.

\begin{acknowledgments}
We are grateful to Naoto Shiraishi and Hal Tasaki for informing us about their results prior to publication.
We especially thank Akihiro Hokkyo for pointing out that the model on the honeycomb lattice can be reduced to the model on the square lattice, as explained in Sec.~\ref{sec:Honeycomb}.
We also thank Akira Shimizu and Ryusuke Hamazaki for their valuable comments on the manuscript.
Y.C. is supported by the Special Postdoctoral Researchers Program at RIKEN and JST ERATO Grant No.~JPMJER2302, Japan.
\end{acknowledgments}

\bibliography{main}

\end{document}